\documentclass[twocolumn]{aastex631}
\usepackage{enumerate}

\usepackage{lineno}

\shorttitle{Photoionization models of the M31 LINER}
\shortauthors{Li et al.}

\usepackage{graphicx}

\defcitealias{2023ApJ...958...89L}{L23}

\begin{document}

\title{Ruling out conventional photoionization models in the closest LINER M31 with CFHT/SITELLE observations}

\author{Zongnan Li}
\email{zongnanli@kasi.re.kr}
\affiliation{Korea Astronomy and Space Science Institute, 776 Daedeok-daero, Yuseong-gu, Daejeon 34055, Republic of Korea}
\affiliation{East Asian Core Observatories Association (EACOA) Fellow}
\author{Zhiyuan Li}
\affiliation{School of Astronomy and Space Science, Nanjing University, Nanjing 210023, China}
\affiliation{Key Laboratory of Modern Astronomy and Astrophysics, Nanjing University, Nanjing 210023, China}
\author{Sumin Wang}
\affiliation{School of Astronomy and Space Science, Nanjing University, Nanjing 210023, China}
\affiliation{Key Laboratory of Modern Astronomy and Astrophysics, Nanjing University, Nanjing 210023, China}
\author{Rub\'{e}n Garc\'{\i}a-Benito}
\affiliation{Instituto de Astrof\'{\i}sica de Andaluc\'{\i}a (CSIC), P.O. Box 3004, 18080 Granada, Spain}
\author{Yifei Jin}
\affiliation{Institute of Natural Sciences, Westlake Institute for Advanced Study, Hangzhou 310024, People's Republic of China}
\affiliation{Department of Astronomy, School of Science, Westlake University, Hangzhou, Zhejiang 310030, People's Republic of China}
\affiliation{Center for Astrophysics, Harvard \& Smithsonian, 60 Garden Street, Cambridge, MA 02138, USA}

\begin{abstract}

The ionization mechanisms of low-ionization nuclear emission-line regions (LINERs), which are common in the local Universe, have been debated for decades. Our nearest large neighbor, M31, is classified as a LINER based on its optical emission-line properties within the central kpc. In this work, we present a detailed photoionization modeling of the circumnuclear ionized gas in M31, explicitly tailored to its well-constrained physical conditions, including the absence of ongoing star formation and a currently inactive active galactic nucleus (AGN). Using spatially resolved CFHT/SITELLE observations, we find that photoionization by hot, evolved low-mass stars distributed throughout the bulge can roughly reproduce the observed radial intensity profiles of H$\alpha$, H$\beta$, and [N\,II]. However, these models fail to match the observed [O\,III] emission, producing radial profiles and [O\,III]/H$\beta$ ratios that are significantly steeper than observed. This discrepancy indicates a deficit of high-energy ionizing photons in standard stellar photoionization models, even with extended ionizing sources. We explore whether this tension can be alleviated by invoking either a bulge-filling, low-density ionized medium surrounding a denser H$\alpha$-emitting disk, or enhanced AGN activity in the recent past. While both scenarios can partially increase the [O\,III] emission, neither provides a fully satisfactory explanation under physically plausible conditions. Together with our earlier results for M81, these findings underscore persistent challenges in explaining LINER-like emission solely through conventional photoionization mechanisms.

\end{abstract}


\section{Introduction} \label{sec:intro}


Low-ionization nuclear emission line regions (LINERs, \citealp{1980A&A....87..152H}) are ubiquitous phenomena in the local universe ($\gtrsim 50\%$ galaxies are LINERs, \citealp{1997ApJ...487..568H}) and have long been considered a lower extension of active galactic nuclei (AGNs). However, their origins remain elusive, as it is relatively easy to produce LINER-like emissions. Although photoionization from low-luminosity AGNs (LLAGNs) is still the leading source of LINERs, other mechanisms, such as hot massive stars \citep{1992ApJ...397L..79F}, X-ray binaries \citep{2006A&A...460...45G}, shocks \citep{1980A&A....87..152H, 1997ApJ...490..202D}, cosmic rays \citep{1984ApJ...286...42F}, and hot low-mass evolved stars (HOLMES; \citealp{1994A&A...292...13B}) are also possible mechanisms for LINERs. In particular, the discovery of extended low-ionization emission-line regions (LIERs; \citealp{2013A&A...558A..43S, 2016MNRAS.461.3111B}) on galaxy scales with the advent of large IFU surveys has made HOLMES the most appealing mechanism for explaining these phenomena. 

However, our previous study (\citealp{2023ApJ...958...89L}, hereafter \citetalias{2023ApJ...958...89L}) found that HOLMES struggled to explain the extended high flux of the [O\,III]$\lambda$5007 line across the bulge of the closest LLAGN, M81 \citep{1999ApJ...516..672H}. 
We constructed detailed photoionization models using {\sc cloudy} \citep{2017RMxAA..53..385F}, with careful treatment of the spatial distribution and SED of ionizing sources (LLAGN and stars), which are well-constrained by multi-band observations \citep{2014MNRAS.438.2804N}. The model-predicted [OIII] line flux is significantly lower than observations beyond 100 pc, where HOLMES is supposed to be the dominant ionizing source. However, the shortcomings of HOLMES in reproducing extended [O III] emission in M81 raise a broader question: is this difficulty unique to M81, or does it reflect a more general limitation of HOLMES-driven photoionization in quiescent galactic nuclei? To address this, a natural next step is to examine M31, a galaxy that closely resembles M81 in its global properties but hosts an even more quiescent nucleus.

In the local universe, our neighbor M31 hosts the nearest LINER \citep{1988AJ.....95..438C}, manifested with prominent [NII]+H$\alpha$ lines across the central kpc, consisting of a spiral-like structure called the \textit{nuclear spiral} \citep{1985ApJ...290..136J}. However, it hosts one of the most quiescent nuclei known, with an Eddington ratio of $10^{-9}$ \citep{2009MNRAS.397..148L}, given its massive black hole ($10^8~\rm M_\odot$, \citealp{2005ApJ...631..280B}). The absence of an AGN \citep{2011ApJ...728L..10L} or recent star formation activity \citep{1998ApJ...504..113B, 2009MNRAS.397..148L, 2015MNRAS.451.4126D} in this nucleus makes HOLMES the most viable mechanism for the LINER emission. M31 shares many similarities with M81. For instance, they are both early-type spiral galaxies, with similar SMBH mass ($\sim 10^8\rm~M_\odot$; \citealp{2003AJ....125.1226D, 2005ApJ...631..280B}), stellar mass and disk size \citep{2011ApJ...739...20C, 2015ApJS..219....5Q}, as well as a low star formation rate and a deficiency of cold gas in the central kpc (e.g. \citealp{2000AJ....119.2745K, 2009MNRAS.399.1026B, 2013A&A...549A..27M, 2015ApJ...805..183L, 2019MNRAS.484..964L}). In addition, they both have a super solar metallicity within the central kpc (1.5-3 $Z_\odot$; \citealp{2000AJ....119.2745K, 2018A&A...618A.156S}). These shared characteristics make M31 an ideal comparison for assessing the viability of HOLMES as the dominant ionizing mechanism in LINER-like nuclei. 

In this work, we extend the photoionization modeling framework developed for M81 to M31 to investigate the ionization mechanism in the nearest LINER. This is the first study to quantify the radial distribution of the major emission lines in M31 and to compare them directly with photoionization models. Despite the spatially distributed nature of the evolved stellar population, we find that HOLMES alone again fails to reproduce the observed ionized gas flux distributions, showing a pronounced deficit in [OIII] emission, mirroring the discrepancy found in M81. The paper is organized as follows. Section \ref{sec:data} describes the CFHT/SITELLE observations, data reduction, and emission line measurements. Section \ref{sec:model} outlines the configuration of the {\sc cloudy} photoionization models. Section \ref{sec:results} presents the model predictions and compares them with the observations. Section \ref{sec:discussion} summarizes our findings and discusses their implications for the ionization mechanism in LINERs.

\section{CFHT/SITELLE observation and data reduction}\label{sec:data}

We utilized integral-field unit (IFU) observations with the SITELLE imaging Fourier transform spectrograph installed on the Canada-France-Hawaii Telescope (CFHT, \citealp{2019MNRAS.485.3930D}). The IFU data were retrieved from the Canadian Astronomical Data Center (CADC\footnote{https://www.cadc-ccda.hia-iha.nrc-cnrc.gc.ca/AdvancedSearch}). The observations (Proposal ID: 16BC19, \citealp{2018MNRAS.473.4130M}) cover an $11' \times 11'$ region centered on the nucleus of M31, using filters SN2 (480--520 nm) and SN3 (647--685 nm) to characterize emission line stellar objects (e.g., planetary nebulae) as well as circumnuclear ionized gas.
The SN2 filter includes H$\beta$ and the [OIII]$\lambda\lambda$4959,5007 doublets, while the SN3 encompasses H$\alpha$ and the [NII]$\lambda\lambda$6548,6584 and [SII]$\lambda\lambda$6716,6731 doublets. 
The observations have a natal pixel of 0.32$''$ and a seeing of $\sim 1''$. The spectral resolution is $R = 5000$ for SN3 (corresponding to a full width at half maximum of $1.3~\text{\AA}$ at the wavelength of H$\alpha$) and $R = 2000$ for SN2 ($2.8~\text{\AA}$ at [OIII]). More details about the observations can be found in \citet{2018MNRAS.473.4130M}. 

\begin{figure*}[hbtp!]
    \centering
    \includegraphics[width=0.35\linewidth]{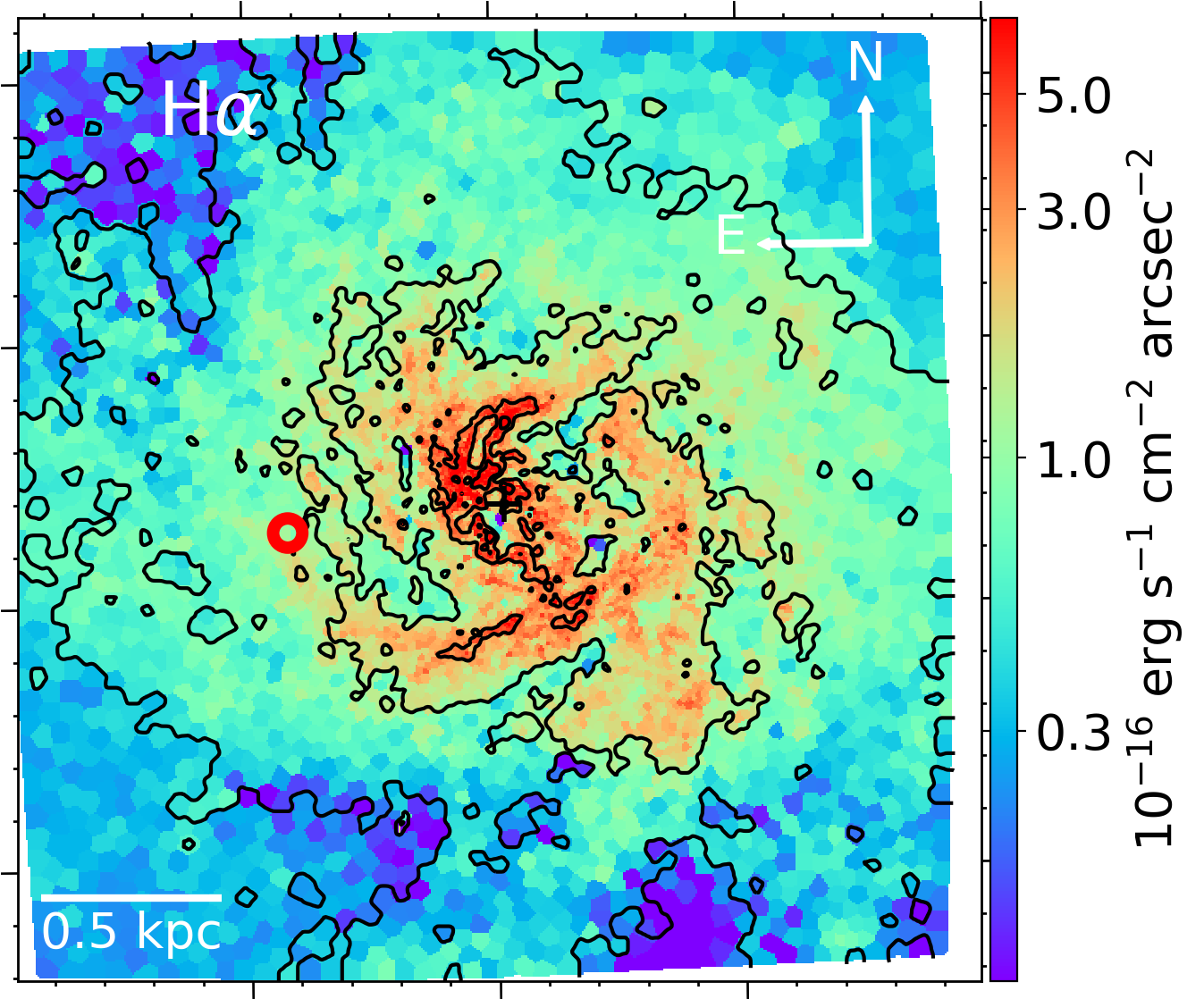}
    \includegraphics[width=0.61\linewidth]{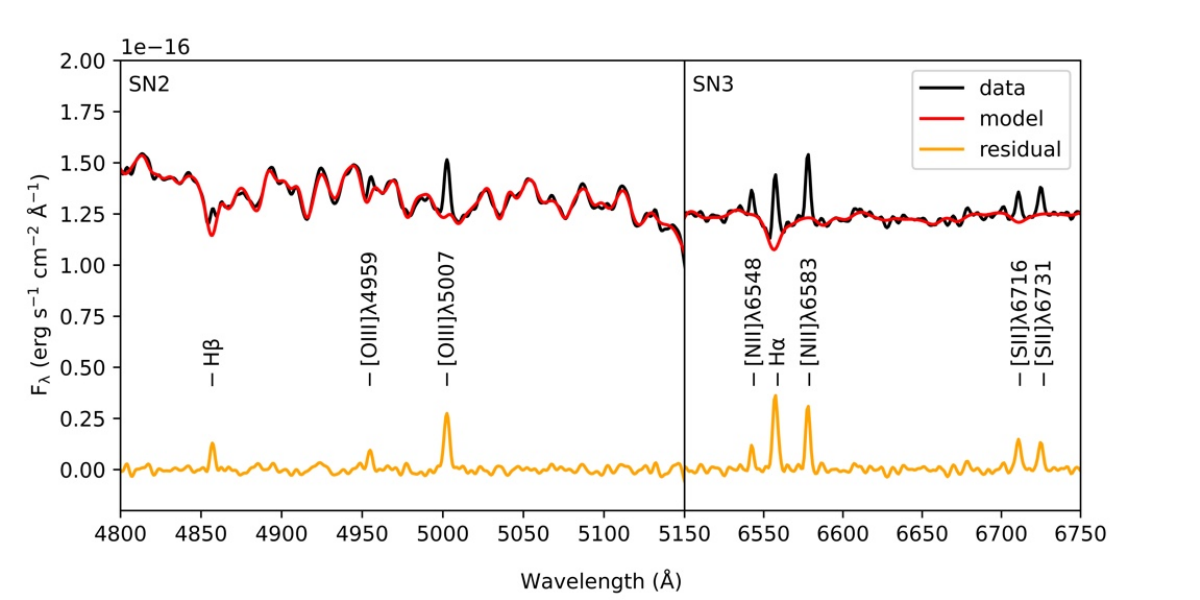}
    \caption{$Left$: H$\alpha$ emission from SITELLE observations, \textbf{overlaid with [OIII] contours with levels 0.3, 1.0, 3.0, 5.0$\times 10^{-16}~\rm erg~s^{-1}~cm^{-2}~arcsec^{-2}$, where the lowest contour corresponds to a $5\sigma$ detection.} $Right:$ Example spectrum extracted from the red circle shown in the left panel. This region lies at the edge of the nuclear spiral and is representative of low surface brightness emission. The original SN2 and SN3 spectra are shown in black, and the stellar continuum model fitted with pPXF is shown in red. After continuum subtraction, the major emission lines become clearly visible in the residual spectrum shown in orange.}
    \label{fig:fov}
\end{figure*}

The data processing followed the standardized pipeline of the SITELLE instrument using the ORBS software \citep{2012SPIE.8451E..3KM, 2015ASPC..495..327M}. The main steps include spatial calibration, wavelength calibration, and flux calibration \citep{2018MNRAS.473.4130M, 2025ApJ...983..182L}. 
The wavelength was calibrated using sky lines described in \citet{2016sf2a.conf...23M}. 
After skyline correction, the line-of-sight velocity was converted to the heliocentric frame. Point sources were also masked from the data \citep{2025ApJ...983..182L}.

To increase the signal-to-noise ratio of the emission lines, Voronoi binning \citep{2003MNRAS.342..345C} was applied on the data cube based on the criteria that the fitted [OIII] emission line has an S/N $>$ 10. After fitting and subtracting the continuum of each bin with the penalized pixel-fixing method (pPXF; \citealp{2004PASP..116..138C}) adopting the E-MILES stellar population templates \citep{2016MNRAS.463.3409V}, the major emission lines are fitted with a single Gaussian to derive the basic properties, including the flux, central velocity, and velocity dispersion. As illustrated by the representative spectrum in Figure \ref{fig:fov}, the majority of emission lines can be adequately modeled by a single Gaussian component. Only about $\sim20\%$ of the pixels show significant double-component profiles \citep{2025ApJ...983..182L}, predominantly located along the edges of the nuclear spiral as reported by \citet{2025A&A...695A.194C}. Given their limited spatial extent and minor impact on global trends, we adopt single Gaussian fits for our analysis of the overall radial distribution of the emission lines, which have a negligible effect on our results and conclusions.
The H$\alpha$ emission exhibits a prominent spiral-like structure across the central kpc, surrounded by an incomplete ring-like structure, hereafter the nuclear spiral (Figure \ref{fig:fov}; \citealp{2025ApJ...983..182L}). The typical H$\alpha$ surface brightness is $\sim10^{-16}~\mathrm{erg~s^{-1}~cm^{-2}~arcsec^{-2}}$, with S/N $> 10$ across most regions of the nuclear spiral. These gaseous features could be explained by the gas spiraling into the central region due to the bar torque after a nearly head-on collision with the compact satellite galaxy M32 \citep{2025ApJ...983..182L}. However, the ionization mechanism of the nuclear spirals remains poorly understood, given the absence of energetic sources in this quiescent nucleus.

\begin{figure*}[hbtp!]
    \centering
    \includegraphics[width=\linewidth]{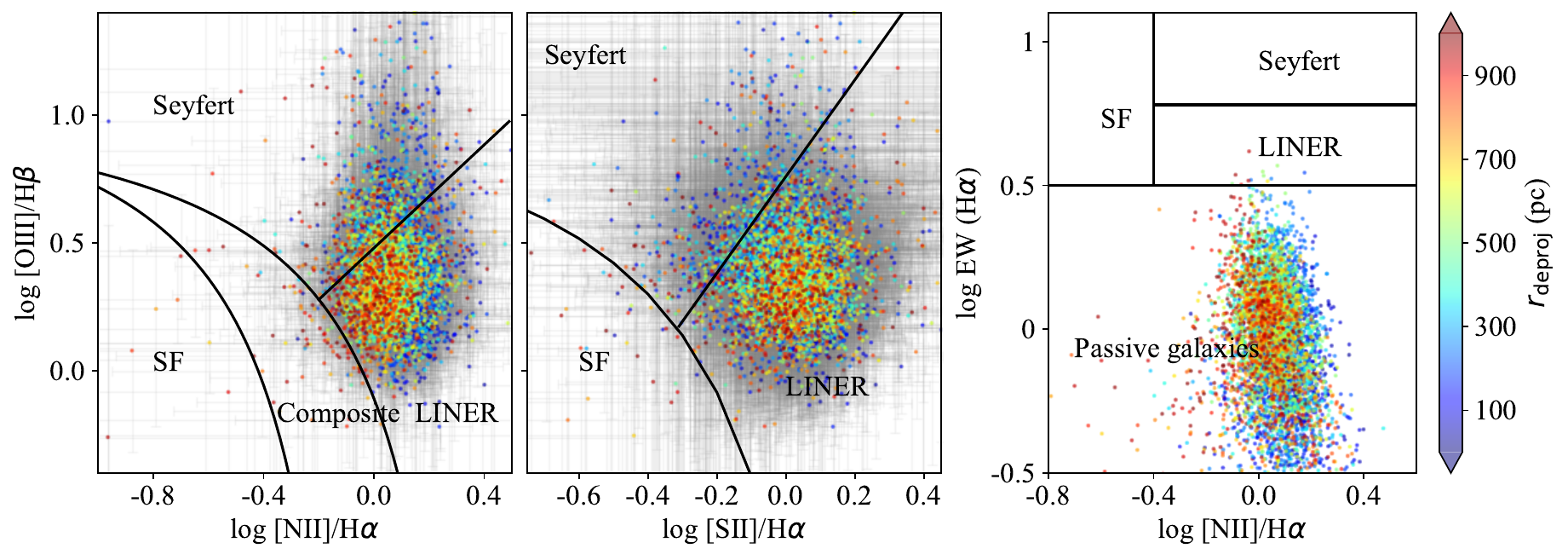}
    \caption{$Left~and~middle:$ Conventional BPT diagrams of the bulge region. 
    The dividing lines are taken from \citet{2006MNRAS.372..961K}, which separate the regions into Seyfert, star formation (SF), LINER, and composite categories, corresponding to the main ionization mechanisms. The [OI]/H$\alpha$ diagram is not shown because the observations do not cover the [OI]$\lambda 6300$ line. Each data point represents a Voronoi bin, and the errorbar represents the statistical uncertainty of the data. The points are color-coded with the deprojected radius from the center. $Right:$ The H$\alpha$ equivalent width versus [NII]/H$\alpha$ (WHAN) diagram \citep{2011MNRAS.413.1687C} of the bulge region. This diagram further separates the traditional LINER region into weak AGN (LINER) and passive galaxies (ionized by HOLMES). }
    \label{fig:bpt}
\end{figure*}

The spatially resolved BPT diagrams \citep{Baldwin_1981, 2006MNRAS.372..961K}, shown in Figure~\ref{fig:bpt}, classify this region primarily as a LINER, albeit with large scatter. The deprojected radius ($r_{\rm deproj}$) of each Voronoi bin is calculated under the assumption that the gas clouds reside in a disk coplanar with the nuclear spiral (see Figure~3 of \citealp{2025ApJ...983..182L}). We adopt an inclination angle of $45^\circ$ for the central region \citep{1988AJ.....95..438C, 2025ApJ...983..182L} and a position angle of $38^\circ$ for the deprojection geometry. Interestingly, signatures characteristic of Seyferts are also present, even out to a radius of $\sim1$ kpc (red points). This indicates that spatially resolved spectroscopy may reveal Seyfert-like features even in the absence of significant AGN activity. A similar phenomenon is seen in the second-nearest LINER, M81, where Seyfert-like emission is detected on kpc scales and cannot be attributed to the central low-luminosity AGN \citepalias{2023ApJ...958...89L}. In contrast, the $W_{\rm H\alpha}$ (H$\alpha$ equivalent width) – [NII]/H$\alpha$ (WHAN) diagram \citep{2011MNRAS.413.1687C} classifies the entire region as a passive galaxy, indicating photoionization dominated by HOLMES. The classification of passive galaxies in the WHAN diagram can be attributed to the low H$\alpha$ EW in the bulge, consistent with the gas-poor nature in this region \citep{2009ApJ...705.1395C}. Taken together, the BPT and WHAN diagnostics suggest that HOLMES should be sufficient to explain the LINER-like emission in M31. To test this, we constructed spatially resolved photoionization models with a detailed treatment of HOLMES, as described below.

\section{Photoionization models}\label{sec:model}

\begin{figure}[htp!]
    \centering
    \includegraphics[width=\linewidth]{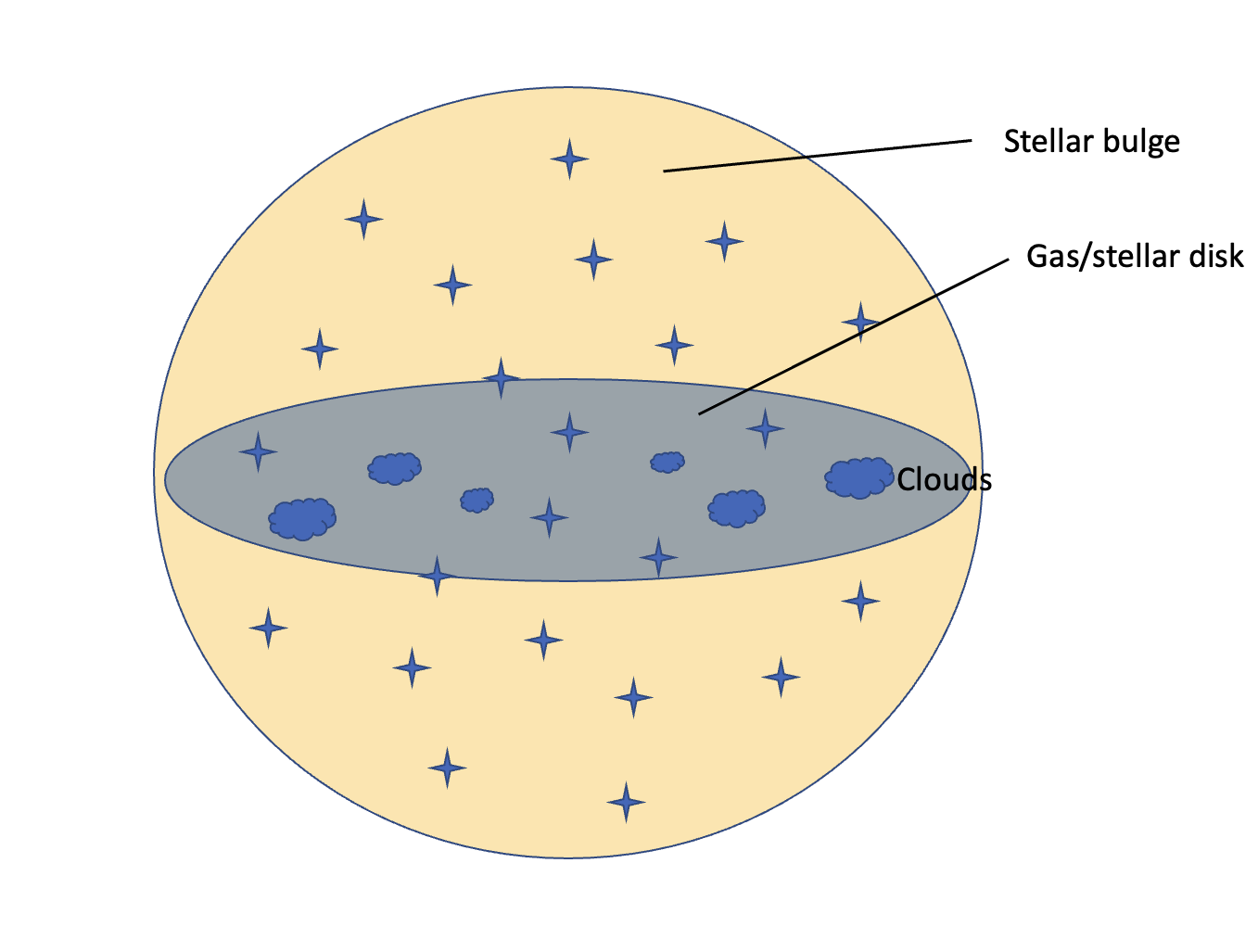}
    \caption{A cartoon of the adopted geometry. Radially spaced clouds are distributed in the plane of the nuclear gas disk (nuclear spiral in M31's case) and are illuminated by the combined radiation field of bulge and disk stars.}
    \label{fig:geo}
\end{figure}

We investigated the ionization mechanism in M31 using photoionization models with the code {\sc cloudy} \citep{2017RMxAA..53..385F, 2023RMxAA..59..327C, 2025arXiv250801102G}, with a focus on testing whether traditional HOLMES-dominated photoionization can explain the spatially extended LINER-like emission, as suggested by the BPT and WHAN diagrams. This code performs a 1D radiative transfer calculation in a gas slab with an input radiation field and a specified density, composition, and geometry. The configuration of the models is similar to that of the M81 bulge \citepalias{2023ApJ...958...89L}, with the main points summarized below:

1. \textit{Geometry}. The line-emitting clouds are assumed to be distributed within a gaseous disk co-planar with the stellar disk. This assumption is supported by the fact that the bulk ionized gas is in the nuclear spiral structure that apparently lies in a disk (hereafter the nuclear gas disk). Minor gas clouds may reside in the bulge region outside the disk, but the contribution from off-plane clouds to the overall emission-line flux has been proven to be negligible \citepalias{2023ApJ...958...89L}. Twenty equally spaced clouds are generated, spanning 10-1000 pc radius, illuminated by stellar populations in both bulge and disk. Open geometry is adopted since the covering factor of the clouds in the nuclear region is supposed to be small, given the low extinction of the nuclear spiral \citep{2016MNRAS.459.2262D}. An illustration of this geometry is presented in Figure \ref{fig:geo}, which closely resembles that proposed for M81 \citepalias{2023ApJ...958...89L}. The primary difference is the absence of an active LLAGN at the center of M31.

\begin{figure}[htp!]
    \centering
    \includegraphics[width=\linewidth]{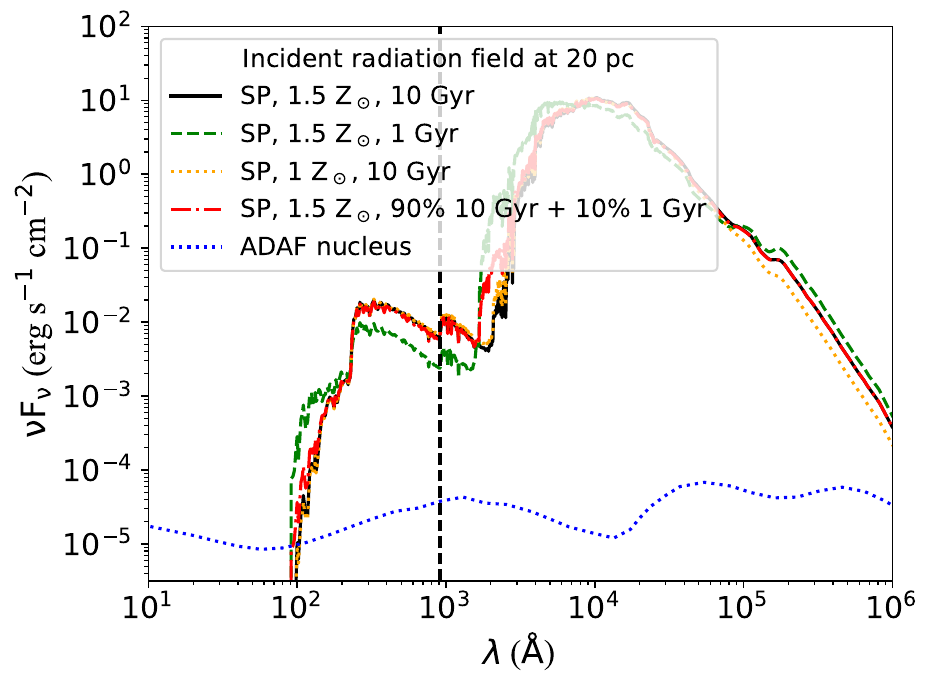}
    \caption{Spectral energy distribution (SED) of M31* (dotted blue curve) and stellar populations with age 10 Gyr and metallicity 1.5 $Z_\odot$ (solid black curve). Additional SP models are shown for (age = 1~Gyr, $Z = 1.5\,Z_\odot$), (age = 10~Gyr, $Z = 2\,Z_\odot$), and a composite population consisting of 90\% 10~Gyr + 10\% 1~Gyr at $Z = 1.5\,Z_\odot$, plotted with green dashed, orange dotted, and red dash-dotted curves, respectively.
    }
    \label{fig:sed}
\end{figure}

2. \textit{Input radiation field.} Given the lack of AGN activity in the M31 center, the ionization photons are mainly from the stellar populations (SPs). The radiation field is modeled as a composite structure: a stellar bulge with a Sersic index of 2.2 and an effective radius of 1 kpc, plus an exponential disk with a scale height of 5.3 kpc. The parameters are adopted from the decomposition of Spitzer/IRAC images \citep{2011ApJ...739...20C}. The SED of stellar populations is generated with Flexible Stellar Population Synthesis (FSPS, \citealp{2009ApJ...699..486C, 2010ApJ...712..833C}), which consists of single-age, single-metallicity templates spanning all available metallicities and ages. The stellar population properties in the bulge, including age, mass, and metallicity, are well constrained by multi-wavelength analyses (e.g., \citealp{2015ApJ...805..183L, 2018A&A...618A.156S}) and, in particular, by the spatially resolved stellar population study from the Panchromatic Hubble Andromeda Treasury (PHAT) survey \citep{2018MNRAS.478.5379D}. These works show that the majority of bulge stars ($>80\%$) are old ($>10$ Gyr) and metal-rich ($>1.5\,Z_\odot$). Although the PHAT star-formation history analysis suggests the presence of a minor younger component ($<10\%$) with an age of $\sim$1 Gyr and solar metallicity \citep{2018MNRAS.478.5379D}, its contribution to the ionizing spectrum is negligible (see red dash-dotted line in Figure \ref{fig:sed}). Hence, we adopt single stellar populations with ages of 10 Gyr and 3 Gyr and metallicities of $1.5\,Z_\odot$ and $Z_\odot$ for the bulge and disk, respectively \citep{2018A&A...618A.156S}. 
A total of $10^{7}$ representing stars are simulated, equally distributed between the bulge and the disk. Their radial positions follow a deprojected Sersic profile \citep{1968adga.book.....S,1997A&A...321..111P} with the adopted geometry parameters \citep{2011ApJ...739...20C}, while the azimuthal ($\phi$) and polar ($\theta$) coordinates are assigned randomly. The luminosity of each simulated star is scaled from the total stellar masses ($M\rm_{bulge} = 3.1 \times 10^{10}~M_\odot$ and $M\rm_{disk} = 5.6 \times 10^{10}~M_\odot$) derived from SED fitting \citep{2012A&A...546A...4T} and a mass-to-light ratio of 0.8 \citep{2018MNRAS.481.3210B}.
The incident radiation field from the SP is then calculated by summing the mean intensity of all stars at the illuminated face of the cloud. 

To evaluate the contribution from the central SMBH M31*, we also compared its SED with that of the SP, as shown in Figure \ref{fig:sed}. 
Since M31* is extremely quiescent, its SED can be described by a radiatively inefficient advection-dominated accretion flow (ADAF; \citealp{1998tbha.conf..148N}), with a bolometric luminosity of $\sim 10^{37}~\rm erg~s^{-1}$ constrained by X-ray observations \citep{2009MNRAS.397..148L}. The corresponding flux is calculated at a distance of 20 pc from the center.
We find that the intensity of M31* is negligible compared to that of the SP, even within the inner 20 pc. Therefore, photoionization from AGN can be safely ignored, and the models discussed below consider only SPs. 

It is well established that the dominant ionizing photons in such environments originate from post-AGB stars in old stellar populations \citep{2011MNRAS.415.2182F}. Indeed, UV-bright stars detected in the inner bulge of M31 by the PHAT survey \citep{2012ApJ...755..131R} are primarily post-AGB and post-early-AGB stars, with no stellar populations younger than 10 Myr identified. M31 remains the only massive bulge in which this UV-bright evolved population has been directly resolved. Notably, for stellar ages $>10^9$ years, the SEDs show only weak dependence on the specific SP age (Figure \ref{fig:sed}). Similarly, SP metallicities have a minor influence on the overall SED within the range constrained by observations \citep{2018A&A...618A.156S, 2018MNRAS.478.5379D}, as illustrated in Figure \ref{fig:sed}. Potential contributions from other stellar components, such as unresolved planetary nebulae (PNe), white dwarfs, and X-ray binaries, are expected to be minor, contributing $\lesssim 10\%$ of the total ionizing photon budget \citepalias{2023ApJ...958...89L}. The contributions from young stars in the disc and old stars in the halo are negligible. The disc is effectively shielded by cold gas and dust, while the halo, with its relatively low stellar mass ($\sim 8\times 10^9~\rm M_\odot$, \citealp{2014ApJ...780..128I}) and large distance from the nuclear spiral, yields only a small ionization parameter. 

3. \textit{Electron density}. The electron density profile is derived from the radial distribution of the [S\textsc{ii}] $\lambda\lambda6716/6731$ line ratio using the calibration of \citet{2014A&A...561A..10P}. The measured $n_{\rm e}$ distribution is approximately flat out to $\sim1$ kpc, with a median value of $\sim300~\rm cm^{-3}$ (Appendix~\ref{sec:a1}). We therefore adopt a constant fiducial electron density of $n_{\rm e}=300~\rm cm^{-3}$ in the default model. 
We note that adopting a density in the range 50-500 cm$^{-3}$ will have little impact on our main conclusion (see Model B1 below as an illustrative example). 
In addition, excluding spatial bins with only upper limits ($n_{\rm e}\le10~\rm cm^{-3}$) does not alter the overall radial trends of the emission-line surface brightness.

4. \textit{Other setup}. The hydrogen column density $N\rm_H$ is assumed to be 10$^{20}~\rm cm^{-2}$, consistent with the low extinction in this region derived from HST observations ($A\rm_V = 0.014 \pm 0.054$; \citealp{2016MNRAS.459.2262D}). The typical cloud thickness is 0.01 pc under this column density, while larger for higher column densities. We note that column densities between 10$^{20-22}~\rm cm^{-2}$ have little influence on our results, indicating that the models are predominantly radiation-bounded. The volume filling factor is assumed to be 0.1, consistent with a few percent inferred for ionized gas in the M31 bulge \citep{2016MNRAS.459.2262D}. We also tested other filling factors (1 and 0.01) and found that they mainly alter the physical thickness of the model clouds, with negligible impact on the resulting line intensities, similar to the case in M81 \citepalias{2023ApJ...958...89L}. A constant pressure mode is assumed to better represent the more realistic situation where the cloud is in total pressure equilibrium with its environment. The cloud metallicity is assumed to be $1.5\,Z_\odot$, reflecting the super-solar metallicity measured in the central region of M31 \citep{2012MNRAS.427.1463Z, 2014ApJ...780..172D}, while using the standard ISM abundance implemented in \textsc{Cloudy}. 

As discussed in \citetalias{2023ApJ...958...89L}, parameters such as the ages and metallicities of the SPs, cloud metallicity, and column density have only minor effects on the resulting emissions within the observational constraints. In contrast, factors such as AGN luminosity and electron density can significantly influence the emergent line intensities. To evaluate their impact, in addition to the default model (\textbf{Model A}) described above, we consider two additional sets of models. 

\textbf{Models B1-2} test whether a low-density gas component can explain the observations. Both models adopt a fixed column density of $N_{\rm H}=10^{20}~\rm cm^{-2}$ as in Model A. \textbf{Model B1 }assumes $n = 50~\rm cm^{-3}$, corresponding to the 16th percentile of the $n_{\rm e}$ distribution, and thus representing the low-density tail of the observed gas. \textbf{Model B2} adopts $n=1~\rm cm^{-3}$ to explore the extreme case of a bulge-filling, very low-density ionized component. With a filling factor of 0.1, this density and column density combination correspond to a characteristic path length of $\sim300$ pc, comparable to the scale height of a galactic disk. This setup is reminiscent of the diffuse ionized gas (DIG) commonly invoked as an extended, low-density ionized component. We note, however, that the physical properties of DIG are poorly constrained, with $n\rm_e$ spanning $\sim0.01$–$1~\mathrm{cm^{-3}}$ and $N\rm_H \sim 10^{18}$–$10^{20}~\mathrm{cm^{-2}}$ for plausible combinations of density, scale height, and filling factor \citep{2016MNRAS.461.3111B, 2020ApJ...897..143K}. Therefore, Model B2 should be regarded as a phenomenological representation of a low-density, extended ionized component, rather than a detailed physical model of galactic DIGs. For clarity, we refer to this component as a \textit{DIG-like} gas hereafter.


\textbf{Models C1-3} explore the possibility of a past high-state AGN phase to alleviate the observed [OIII] deficiency by injecting more energetic photons. All three models adopt the same underlying stellar population as \textbf{Model A}. In Model C1, the AGN luminosity is set to $10^{-6}\,L_{\rm Edd}$, approximately three orders of magnitude higher than the current level, motivated by Chandra observations of a recent X-ray flare with an inferred jet power of $1.6\times10^{40}\,\mathrm{erg\,s^{-1}}$ \citep{2011ApJ...728L..10L}. Models C2 and C3 assume progressively higher AGN luminosities of $10^{-4}\,L_{\rm Edd}$ and $10^{-2}\,L_{\rm Edd}$, respectively, to assess the level of AGN luminosity required to reproduce the observed [OIII] emission. These luminosities remain within the range of LLAGN ($L_{\rm bol}/L_{\rm Edd} \lesssim 10^{-2}$), for which an ADAF SED is still applicable \citep{2009ApJ...699..626H, 2014MNRAS.438.2804N}.

\section{Results}\label{sec:results}

\begin{figure*}[htp!]
    \centering
    \includegraphics[width=\linewidth]{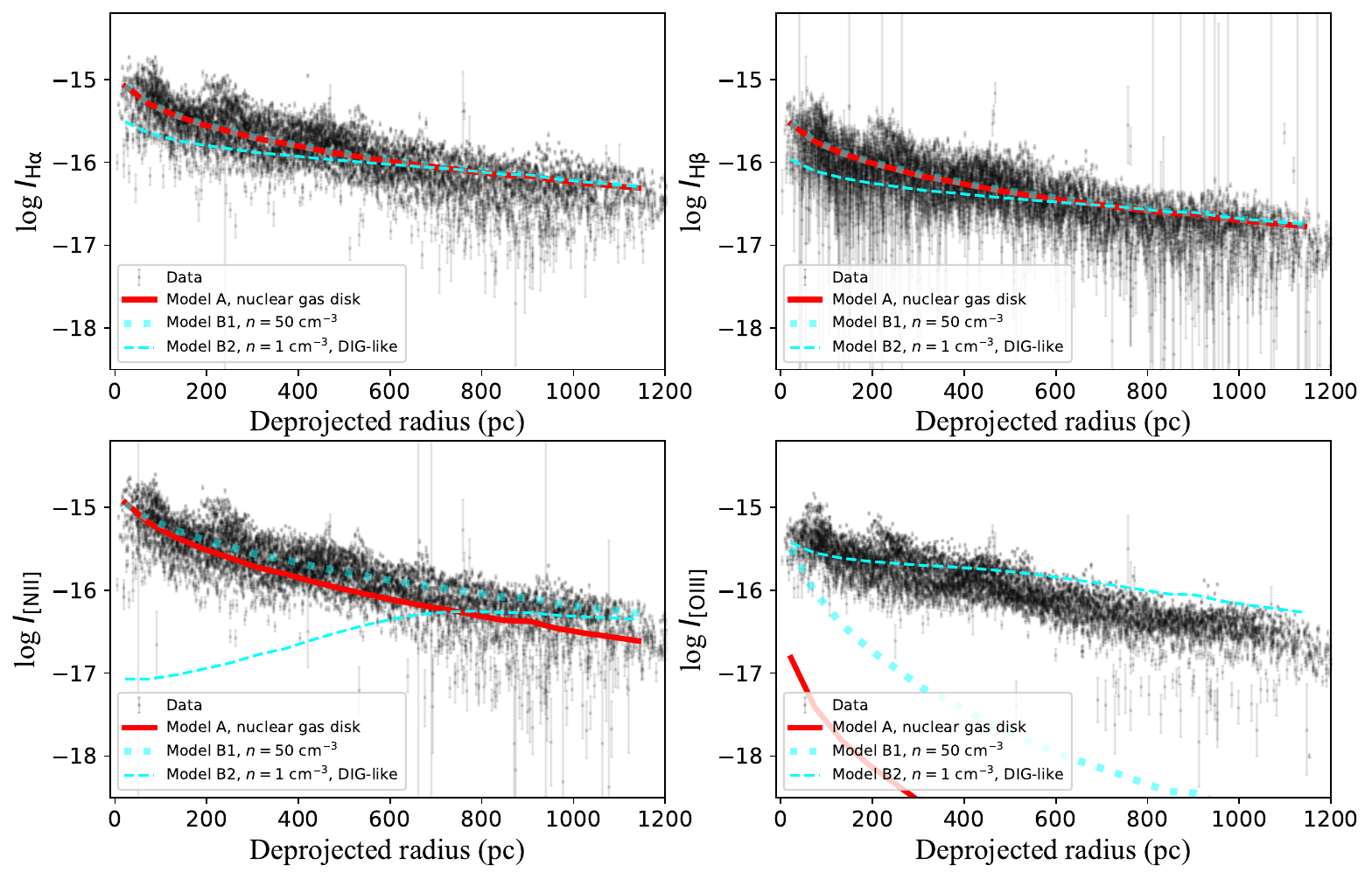}
    \caption{The radial distribution of line intensities of the four major emission lines H$\alpha$, H$\beta$, [OIII], and [NII] compared with CLOUDY models B1-2. Data extracted from each Voronoi bin is shown in black points, while the CLOUDY model A with a constant density representative of the nuclear gas disk as described above, is shown with a red solid line. Model B1 and B2 with lower density of $n= 50~\rm cm^{-3}$ and $n=1~\rm cm^{-3}$ are shown with cyan dotted and dashed lines, respectively.}
    \label{fig:eml1}
\end{figure*}

\begin{figure*}[htp!]
    \centering
    \includegraphics[width=\linewidth]{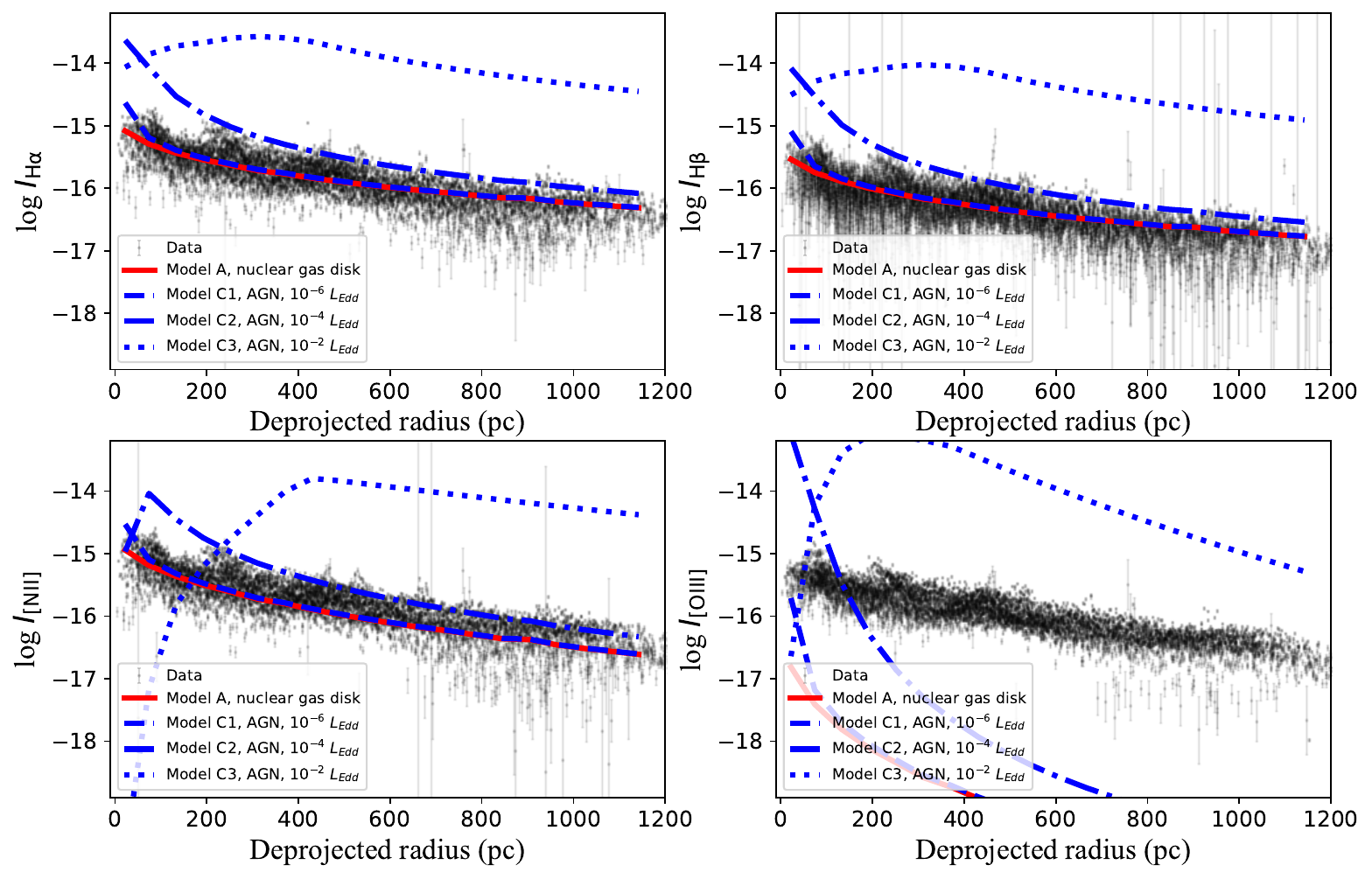}
    \caption{The radial distribution of line intensities of the four major emission lines compared with CLOUDY models C1-3, the symbols are the same as in Figure \ref{fig:eml1}. Models C1-3 with previously enhanced AGN activity of $10^{-6}, ~10^{-4}, 10^{-2}~L\rm_{Edd}$ are shown with blue dashed, dash-dotted, and dotted lines, respectively.}
    \label{fig:eml2}
\end{figure*}

To enable a direct comparison between the observed line intensities and the model predictions, we first corrected the observed fluxes for foreground extinction using the mean bulge value of $A_V = 0.014$ \citep{2016MNRAS.459.2262D}.
The radial distribution of the emission line intensity from the models in comparison with the data is shown in Figures \ref{fig:eml1} and \ref{fig:eml2}. The major emission lines, H$\alpha$, H$\beta$, [NII], and [OIII], exhibit nearly flat radial profiles up to 1 kpc. This contrasts with that of M81, which has an upturn in the central 100 pc, mainly attributable to its LLAGN. The default model, Model A, reproduces the radial profiles of H$\alpha$ and H$\beta$ well, with $\chi^2< 50$, but slightly underpredicts [NII] ($\chi^2 \sim 100$) and significantly underpredicts [OIII] ($\chi^2 > 10^4$)), especially in the outer regions, where the modeled intensity falls over two orders of magnitude below the observed [OIII]. We note that the [N\,\textsc{ii}] emission line is particularly sensitive to the nitrogen abundance, and a modest adjustment of this parameter improves the overall agreement between the models and the observations (see Appendix~\ref{sec:a2} for a fine-tuning model.)

For Model B1 with $n\rm_e = 50~cm^{-3}$, although this model offers a slightly better match to the data than the default Model~A, it still underpredicts [O\,\textsc{iii}] by more than 2~dex at radii beyond 1~kpc (Figure~\ref{fig:eml1}). In Model B2, the DIG-like component embedded in the bulge largely accounts for the high [OIII] flux, particularly in the outer regions, although with a slightly flatter radial trend. This results in a markedly improved fit, with the $\chi^2$ value reduced to 150 compared to $\sim 1.6\times 10^4$ for Model A (see the full $\chi^2$ comparison in Appendix~\ref{sec:a3}).In fact, the DIG-like gas contributes over 95\% of the [OIII] emission at all radii. By contrast, its contribution to the Balmer lines is $\lesssim50\%$, and even smaller to [N\,\textsc{ii}] (only a few percent within $\sim500$~pc and $< 50\%$ beyond), resulting in only a modest impact on their radial distributions.

Model C1 demonstrates that an observation-motivated, moderately elevated AGN state enhances emission lines only within the central 200 pc, with negligible impact at larger radii, and thus fails to reproduce the observed flat radial distribution of [OIII]. Model C2, with an AGN luminosity increased by two orders of magnitude ($10^{-4}\,L_{\rm Edd}$) than Model C1, significantly overpredicts [O \textsc{iii}] in the inner $\sim$200 pc while still underpredicting the emission at larger radii. This model also overpredicts the Balmer lines and produces a central deficit of [N \textsc{ii}] within $\sim$100 pc, where nitrogen is ionized to higher stages by the intense radiation field. The steep radial decline of [OIII] in this model is expected, as the AGN acts as a point source and thus generates a centrally concentrated ionizing field. Model C3, with $L_{\rm AGN}=10^{-2}\,L_{\rm Edd}$, overpredicts all emission lines by $\gtrsim 2~\rm dex$ beyond $\sim$400 pc, while simultaneously producing a decline in the central regions due to over-ionization. Although a sufficiently luminous past AGN can reproduce the high [O\textsc{iii}] surface brightness at large radii (as in Model C3), all such models simultaneously significantly overpredict the Balmer line emission. This tension persists across the explored AGN luminosity range.

\begin{figure*}[htp!]
    \centering
    \includegraphics[width=\linewidth]{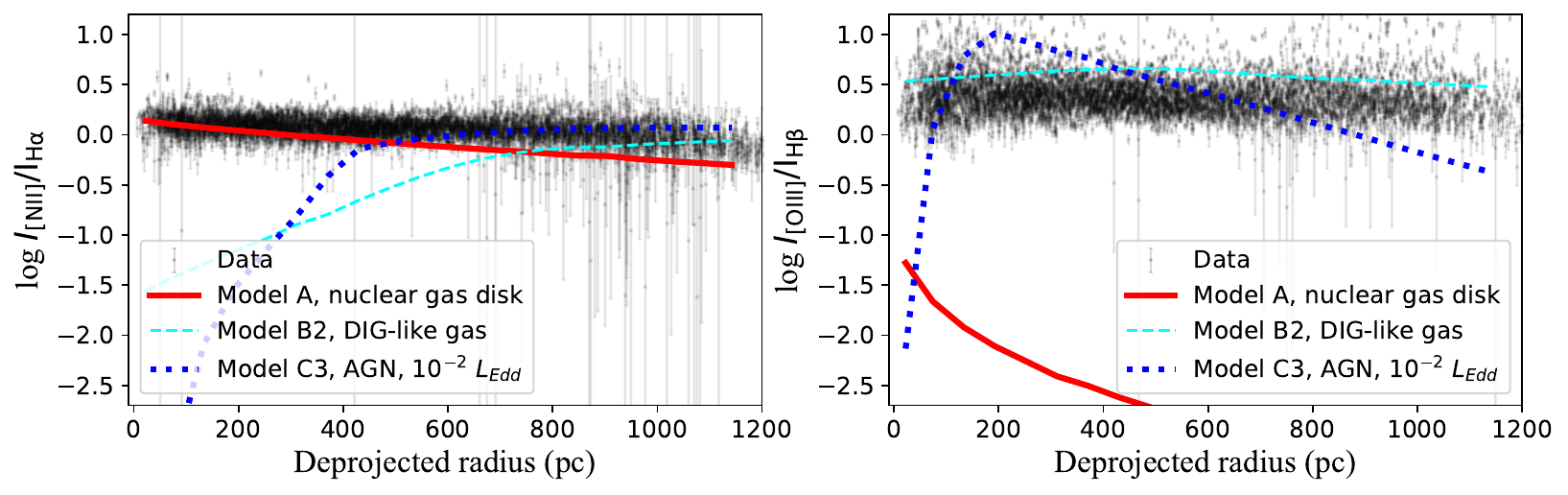}
    \caption{The radial distribution of the emission line ratios compared with the two models that most effectively alleviate the [OIII] deficiency, Models B2 and C3. The symbols are the same as in Figures \ref{fig:eml1} and \ref{fig:eml2}. }
    \label{fig:ratio}
\end{figure*}

Figure~\ref{fig:ratio} shows the radial distribution of [NII]/H$\alpha$ and [OIII]/H$\beta$ line ratios in comparison with the two models that most effectively alleviate the [OIII] deficiency, Model B2 and C3. The line ratios exhibit a remarkable flat radial profile out to 1 kpc, with median values of 1.1 and 2.3 for [NII]/H$\alpha$ and [OIII]/H$\beta$, respectively. In Model A, the predicted [NII]/H$\alpha$ generally matches the observed flat profile, with only a modest $\sim$0.3 dex decline in the outer regions. In contrast, the modeled [OIII]/H$\beta$ remains over two orders of magnitude below the observations. Model B2, with a low density, DIG-like component, successfully reproduces the observed flat [OIII]/H$\beta$ profile, while systematically higher by a factor of $\sim 1.5$. In addition, it shows a low [NII]/H$\alpha$ (log [NII]/H$\alpha < -0.5$) within 500 pc (see discussion in Section \ref{sec:discussion}). Model C3, with a substantially higher AGN luminosity, raises the [O\textsc{iii}]/H$\beta$ ratio to a level comparable to the observations. However, it predicts a radial decline beyond $\sim$200 pc and a central dip within the inner $\sim$200 pc, distinct from the observed flat profile. In addition, the model significantly underpredicts [N \textsc{ii}]/H$\alpha$ within $\sim$400 pc, while providing a reasonable match at larger radii. Taken together, these results indicate that none of the models considered here can simultaneously reproduce the observed flat radial distributions of both line ratios.

\section{Discussion and summary}\label{sec:discussion}

Although the extended emission-line region in M31 was identified decades ago \citep{1985ApJ...290..136J}, our work provides the first quantitative measurement of the radial profiles of its major emission lines and the first direct comparison with physically motivated CLOUDY photoionization models. A key novelty of our modeling is that all primary inputs, including stellar population age, metallicity, spatial distribution, and ionizing SED, are empirically well constrained, reflecting the well-established properties of M31's old, massive bulge. We find that the major emission lines decline smoothly by roughly 1 dex over the central kiloparsec. Unlike M81, which hosts a low-luminosity AGN, M31 shows no central upturn within the inner 100 pc, reinforcing the idea that LLAGNs play a significant role in ionizing circumnuclear gas when present.
As discussed in \citetalias{2023ApJ...958...89L}, variations in stellar population age, metallicity, filling factor, or AGN SED shape within the observational constraints have only a negligible impact on the predicted emission-line strengths, and none can account for the large discrepancy between the observed and modeled [O\,III] flux. In the absence of an active nucleus, old stellar populations underproduce [O\,III] by more than two orders of magnitude even in the innermost region. This behavior mirrors the case in M81, where photoionization by old stars likewise falls short of the observed [O\,III] flux by orders of magnitude beyond $\sim 200$ pc \citepalias{2023ApJ...958...89L}.

The primary cause of this [OIII] deficit is the intrinsically low ionization parameter in the bulge of M31. Defined as $U = \frac{\Phi(\mathrm{H})}{n(\mathrm{H})\,c}$ \citep{1989agna.book.....O}, where $\Phi(\mathrm{H})$ is the ionizing photon flux, $n(\mathrm{H})$ the hydrogen density, and $c$ the speed of light, the ionization parameter depends sensitively on both the intensity of the radiation field and the gas density. In our models, $\log U$ declines smoothly from $-4.5$ at 20 pc to $-5.5$ at 1 kpc, consistent with the observed radial decrease in H$\alpha$, H$\beta$, and [N\,II] fluxes ($\sim$1 dex across the central kpc), resulting in a nearly constant [NII]/H$\alpha$ ratio. However, at such a low $U$, the [O\,III] flux drops much more rapidly than H$\beta$, producing a strongly decreasing [O\,III]/H$\beta$ ratio. This is in stark contrast to high-$U$ conditions ($\log U \gtrsim -3$), where [O\,III]/H$\beta$ remains approximately constant \citep{2008MNRAS.391L..29S, 2017MNRAS.466.3217Z}.
The SED shape and the spatial distribution of the radiation field are well constrained observationally \citep{2011ApJ...739...20C}, and $n_{\rm e}$ can, in principle, be constrained from the [S\,II] doublet using the SITELLE data. Nevertheless, the electron density remains a major source of uncertainty. The [S\,II] ratio is sensitive only to densities in the range $10^{1{-}4}~\mathrm{cm^{-3}}$ (corresponding to ratios 0.45--1.42), and even with careful treatment of uncertainties and upper limits (Appendix~\ref{sec:a1}), the derived densities are not tightly constrained, although the value is consistent with previous findings in nearby galactic nuclei ($n_{\rm e} \sim 30$--$300~\mathrm{cm^{-3}}$; \citealp{2016ApJ...826..175H, 2025arXiv250513677B}). To assess the impact of this uncertainty, we also tested a lower density case of $n_{\rm e} = 50~\mathrm{cm^{-3}}$ (Model B1), but the predicted [O\,III]/H$\beta$ ratio remains $\sim 2$ dex below the observed value at 1 kpc. 
It is possible that the [O\,III]-emitting gas occupies regions with densities different from those probed by [S\,II]. For example, [O\,II]-based density estimates are found to be higher than those from [S\,II] in nearby galaxies \citep{2025arXiv250513677B}. Fine-tuning $n_{\rm e}$ could, in principle, reproduce the observed [O\,III] trend (see discussion of DIG below). However, given the current observational constraints, traditional photoionization models fail to explain the flat radial profiles of [O\,III], despite successfully reproducing the other major emission lines. Quantitatively, reconciling Model A with the observed [O\,III] flux would require either extremely low gas densities ($n_{\rm e} \lesssim 1~\mathrm{cm^{-3}}$) in the [O\,III]-emitting phase, far below densities indicated by the [S\,II] ratio, or an increase in the ionizing photon flux by $\sim$2 dex, which is incompatible with the observed stellar SED and the absence of an AGN. This highlights the difficulty of producing the observed [O\,III] emission with standard photoionization in the bulge of M31.

DIG ($T_{\rm e}\sim10^{4}$ K, $n_{\rm e}\sim10^{-1}$ cm$^{-3}$; \citealp{2016MNRAS.461.3111B}) has been proposed as a potential ionization source in nearby LINERs. However, spatially resolved studies of SDSS MaNGA galaxies indicate that DIG generally contributes only a small fraction of the total [O\,\textsc{iii}] emission, although the [O\,\textsc{iii}]/H$\beta$ ratio can vary with environment \citep{2017A&A...599A.141J, 2017MNRAS.466.3217Z, 2022A&A...659A..26B}. In the disk of M31, DIG is characterized by weak [O\,\textsc{iii}] emission and relatively low [O\,\textsc{iii}]/H$\beta$ ratios, with typical values of $\sim0.5$ \citep{1997ApJ...483..666G}, compared to $\sim2$ measured in the bulge region analyzed here. More generally, DIG is found to exhibit lower ionization parameters than classical H\,\textsc{ii} regions, as inferred from the [O\,\textsc{iii}]/[O\,\textsc{ii}] ratio \citep{2017MNRAS.466.3217Z}.
In contrast, the DIG-like component in Model B2 requires relatively high ionization parameters ($\log U\approx-3$ to $-2$, or even higher for lower density), comparable to or exceeding those of HII regions ($\log U\approx-3.5$ to $-2.5$; \citealp{2019ARA&A..57..511K}), in order to reproduce the observed [O III] emission. This places the model in tension with the typical ionization conditions inferred for DIG in both M31 and other nearby galaxies \citep{2017MNRAS.466.3217Z}, suggesting that while the DIG-like component (Model B2) can reproduce the observed line ratios in a phenomenological sense, it may not correspond to the physical properties of DIG inferred from observations.
Additionally, the cospatiality of H$\alpha$ and [OIII] seems to disfavor the bulge-filling, DIG-like model (Figure \ref{fig:fov}). 
If DIG dominated the [O\,III] emission, as in Model B2, the [O\,III] emission would be expected to extend well beyond the nuclear spiral and deviate significantly from the H$\alpha$ morphology. Instead, [O\,III] closely follows H$\alpha$ and shows no correlation with electron density (Figure~\ref{fig:ne}).
Even if the DIG follows a nuclear spiral-like morphology akin to the nuclear gas disk, their kinematic signatures should differ. Gas associated with H\,II regions in the nuclear disk is expected to have relatively small velocity dispersions, whereas extraplanar diffuse gas typically shows larger dispersions ($\sim10$–$50~\mathrm{km~s^{-1}}$) and a systematic velocity lag relative to the disk gas of $\sim10$–$30~\mathrm{km~s^{-1}}$, as observed in nearby disk galaxies \citep{2017ApJ...839...87B, 2017ApJ...845..155B, 2019ApJ...885..160B, 2020MNRAS.491.4089D, 2021MNRAS.504.3013L}. While such differences are beyond the resolving power of our SITELLE data, they are accessible to the VIRUS-W IFU ($R\sim9000$, $\sigma\sim14$ km s$^{-1}$; \citealp{2018A&A...611A..38O}), which mapped the M31 bulge and inner disk in [O\,III] and H$\beta$. Indeed, VIRUS-W observations reveal two kinematic components in both [O\,III] and H$\beta$ out to radii of $\sim5$ kpc \citep{2018A&A...611A..38O}. However, these components exhibit comparable velocity dispersions ($< 5~\mathrm{km~s^{-1}}$) and fluxes and are separated in velocity by $\sim100~\mathrm{km~s^{-1}}$, a pattern more consistent with large-scale bulk motions than with a superposition of DIG and nuclear disk emission. Nonetheless, higher spectral resolution observations will be required to further assess the viability of the DIG scenario.

Regarding past AGN activity, the situation is similar to that discussed in \citetalias{2023ApJ...958...89L}. AGN ionization is typically highly centrally concentrated, producing a centrally peaked emission-line distribution. Although a sufficiently luminous past AGN outburst could, in principle, elevate the [O \textsc{iii}] emission at large radii (e.g., Model C3), it would simultaneously generate substantially stronger H$\alpha$ emission. Because H$\alpha$ recombines more slowly than [O\textsc{iii}], with characteristic timescales of $\sim10^{4}$ yr versus $\sim10^{3}$ yr under typical conditions ($n \sim 10~\mathrm{cm^{-3}}$, $T \sim 10^{4}$ K; case B recombination; \citealp{1989agna.book.....O}), the H$\alpha$ emission would remain elevated long after the AGN faded, inconsistent with the observed line strengths. A quenched AGN therefore cannot simultaneously reproduce the present-day H$\alpha$ and [O \textsc{iii}] emission at comparable levels.
Even a variable AGN cannot explain the observed flat profile at large radii: the light travel time across 1 kpc is $\sim$3000 years, while [O\,III] decays rapidly at this $n_{\rm e}$, generating a delayed ionization front rather than a flat distribution. More generally, AGN-powered emission is intrinsically centrally concentrated and therefore expected to exhibit a strong radial dependence. As a result, both steady and time-variable AGN activity have difficulty reproducing the nearly flat radial profiles of all emission lines extending out to $\sim$1\,kpc in M31, consistent with earlier findings in M81 \citetalias{2023ApJ...958...89L}. While a comprehensive exploration of AGN SED shapes, variability histories, and duty cycles is beyond the scope of this work, our results suggest that AGN photoionization is unlikely to be the dominant mechanism in these systems. Instead, our findings point to the possible importance of other spatially extended and ubiquitous processes in powering LINERs, such as shocks \citep{2018MNRAS.473.4130M}, cosmic rays \citep{1984ApJ...286...42F}, and radiative turbulent mixing layers \citep{2019MNRAS.487..737J}. A quantitative investigation of these mechanisms is deferred to future work.


In summary, our photoionization models, compared with CFHT/SITELLE IFU data of M31's bulge, reproduce Balmer and [NII] emission lines but fail to account for the flat [OIII] distribution. 
Building upon our earlier findings in M81, the results from M31 strongly suggest that the standard photoionization framework (AGN and HOLMES) is insufficient to reproduce the extended [OIII] emission in quiescent galactic nuclei, under the reasonable assumption that the bulk of the circumnuclear ionized gas resides in a thin plane associated with the nuclear disk. Taken together, these two nearest LINERs point to a more general limitation of HOLMES or LLAGN-driven ionization and underscore the need to identify additional mechanisms responsible for powering LINERs in galaxies. A detailed exploration of such mechanisms is deferred to future work.


\begin{acknowledgments}
This work was supported by the National Key Research and Development Program of China (NO.2022YFF0503402 and No. 2022YFA1605000). Z.N.L. acknowledges support from the China National Postdoctoral Program for Innovation Talents (grant BX20220301) and the East Asian Core Observatories Association Fellowship. R.G.B acknowledges financial support from the Severo Ochoa grant CEX2021-001131-S funded by MCIN AEI/10.13039/501100011033 and PID2022-141755NB-I00.

\end{acknowledgments}

\vspace{5mm}
\facilities{CFHT/SITELLE}


\software{astropy \citep{2013A&A...558A..33A,2018AJ....156..123A},  
          Cloudy \citep{2017RMxAA..53..385F}, ORBS \citep{2012SPIE.8451E..3KM, 2015ASPC..495..327M}
          }



\appendix
\section{Electron density}\label{sec:a1}

The electron density map derived from the [S II] line ratio is shown in Figure~\ref{fig:ne}, overlaid with [O III] intensity contours. The inferred electron density distribution appears spatially irregular, with no clear large-scale pattern, and shows no obvious correlation with the [O III] emission morphology.
The radial distribution of $n\rm_e$ is shown in the right panel of Figure \ref{fig:ne}. The density measurements exhibit substantial scatter across the bulge, with $\sim 40\%$ of spaxels yielding only an upper limit of $\sim10~\rm cm^{-3}$. To account for uncertainties and censored data (upper and lower limits), we performed a linear regression with censoring. The resulting fit converges to a nearly flat profile with a characteristic value of $\sim300~\rm cm^{-3}$. 

\renewcommand{\thefigure}{A\arabic{figure}}
\setcounter{figure}{0}


\begin{figure*}
    \centering
    \includegraphics[width=0.41\linewidth]{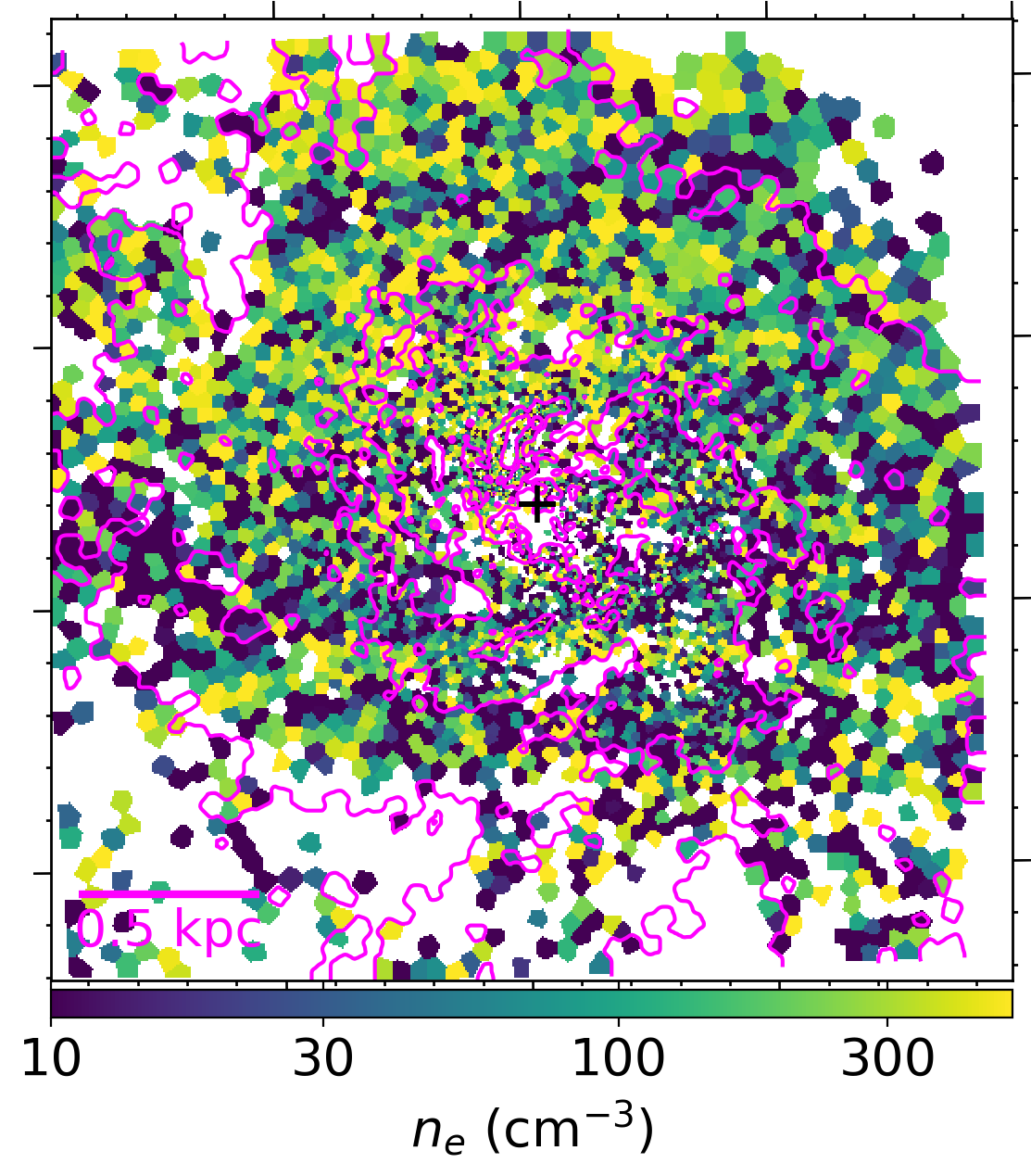}
    \includegraphics[width=0.58\linewidth]{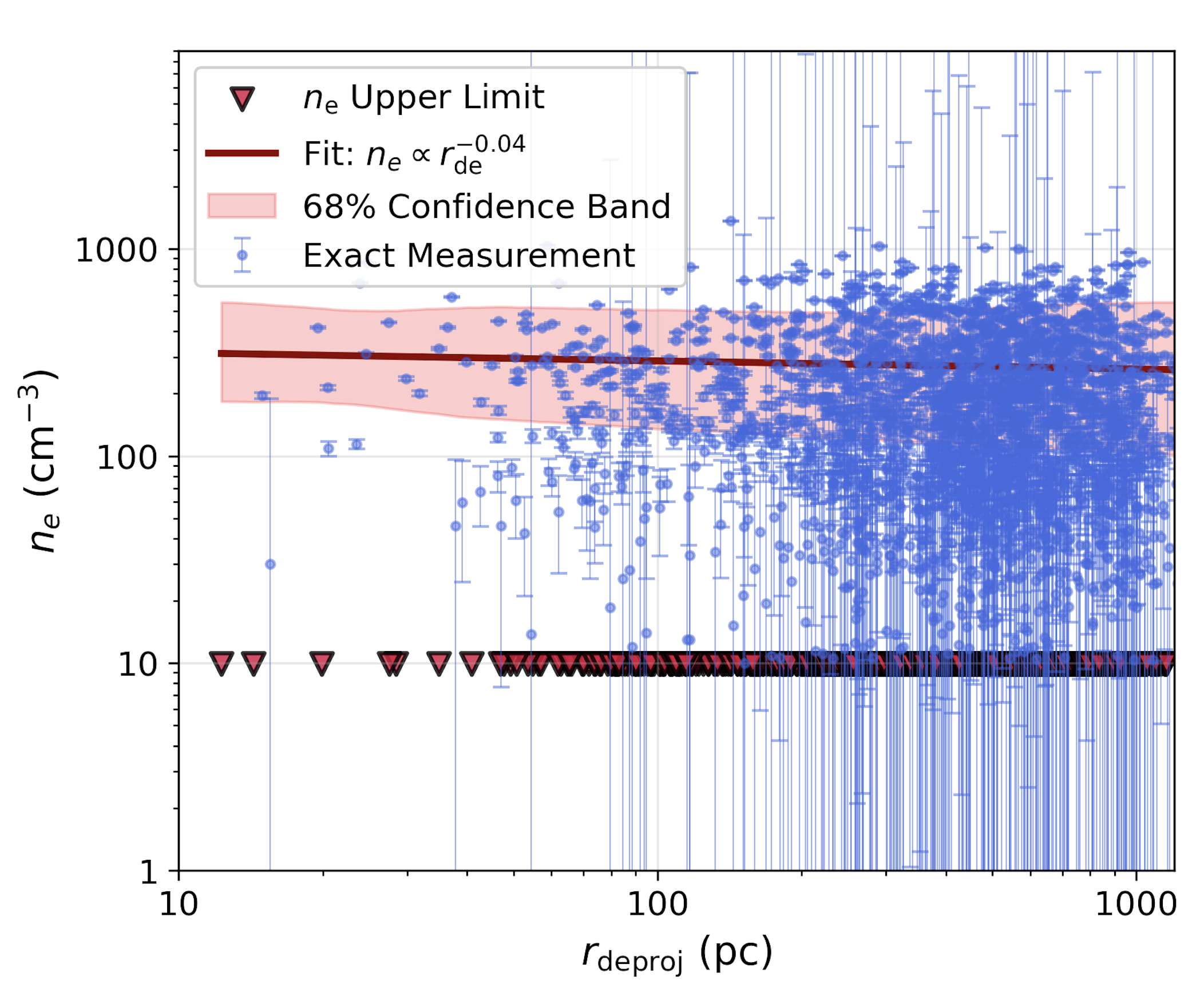}
    \caption{
    $Left:$ Electron density map derived from the [SII] ratio \citep{2014A&A...561A..10P} overlaid with [OIII] contours. The levels are the same as in Figure \ref{fig:fov}. Pixels with [SII] ratio SNR$< 3$ are excluded.
    $Right:$ The radial distribution of $n\rm_e$. The radius is deprojected adopting the common geometric factors \citep{1988AJ.....95..438C}. 
    The $n_{\rm e}$-sensitive range of 10$^{1-4}\rm~cm^{-3}$ (corresponding to a [SII] ratio of 0.45–1.42) is shown as circles and treated as reliable measurements. Values falling outside this range are treated as upper limits (no lower limits are derived) and are displayed as triangles.
    The best-fit power law is shown with a crimson curve with a 68\% confidence interval in red shaded. All uncertainties and limits are incorporated into the linear regression using censoring. 
    }
    \label{fig:ne}
\end{figure*}

\section{Other models}\label{sec:a2}

\begin{figure*}[htp!]
    \centering
    \includegraphics[width=\linewidth]{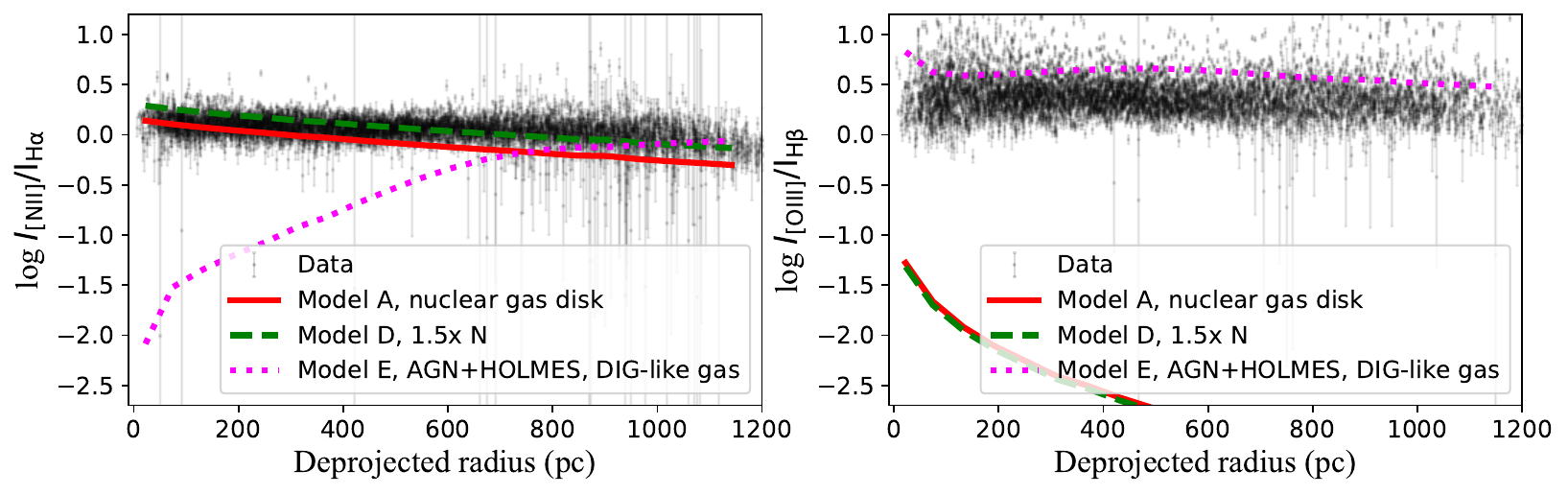}
\caption{The radial distribution of the emission line intensities compared with additional CLOUDY models. The symbols follow the same convention as in Figure~\ref{fig:eml1}. Each additional model varies a single parameter relative to Model~A: model D adopts an N abundance 1.5 times higher with respect to the standard ISM abundance (green dashed line), and model E assumes an AGN + HOLMES as the ionizing source for DIG-like gas (magenta dotted line).}
    \label{fig:other}
\end{figure*}

For completeness, we explored two additional photoionization models and compared the resulting line ratios with observations (Figure \ref{fig:other}): (i) a metallicity fine-tuning model (Model~D), and (ii) a composite model combining AGN and HOLMES ionization for a DIG-like component (Model~E). These models are not discussed in the main text because they produce only modest changes in the predicted emission-line intensities and do not affect our overall conclusions. We briefly summarize them here for reference.

\textbf{Model D: 1.5 times higher N abundance.} Since N is sensitive to metallicity, we included a model with slightly enhanced single element abundance to 1.5 times to account for the underprediction of Model A to [NII]. As shown in Figure \ref{fig:other}, the fine-tuning model provides a more satisfactory fit with a $\chi^2 \sim 60$ compared with $> 100$ in Model A, demonstrating that fine-tuning can improve the agreement between metal lines and models by a factor of a few. One might ask whether variations in gas metallicity could enhance the [O\,III] emission as well and explain the [OIII] deficiency. The bulge ISM of M31 is observed to be supersolar, reaching $\gtrsim 2\,Z_\odot$ within the central kiloparsec \citep{2014ApJ...780..172D} and up to $\sim 3,Z_\odot$ in the central $\sim100$ pc. However, increasing the overall metallicity primarily enhances metal-line cooling, which lowers the electron temperature and effective ionization parameter. As a consequence, collisionally excited lines such as [OIII] are suppressed relative to recombination lines, rather than enhanced. Therefore, variations in global metallicity cannot account for the observed excess [OIII] emission at large radii. Similarly, selectively enhancing the oxygen abundance (by at most a factor of a few within observational constraints) cannot reconcile the $>2$ dex discrepancy between the model predictions and the data.


\textbf{Model E: AGN + HOLMES as the ionizing source for DIG-like gas.}
We further test a model in which the DIG-like component is ionized by a combination of HOLMES and an AGN, adopting the same AGN luminosity as in Model~C1, to evaluate the effect of a moderately enhanced AGN on low-density gas. Relative to Model~B2, the inclusion of an AGN slightly increases [O\,\textsc{iii}]/H$\beta$ and decreases [N\,\textsc{ii}]/H$\alpha$ within the central $\sim$100~pc, consistent with the higher ionization parameter produced by the harder radiation field. Increasing the AGN luminosity leads to qualitatively similar trends as in the dense nuclear gas disk, but with systematically higher ionization parameters owing to the lower gas density: [N\,\textsc{ii}] is further suppressed by over-ionization of N$^{+}$, while [O\,\textsc{iii}] increases until O$^{++}$ is ionized to higher states, at which point the [O\,\textsc{iii}] emission saturates or declines. In either case, AGN-produced emission will have a strong radial dependence, which will have difficulty in explaining the flat distribution.

\section{$\chi^2$ test}\label{sec:a3}

\begin{figure*}[htp!]
    \centering
    \includegraphics[width=0.49\linewidth]{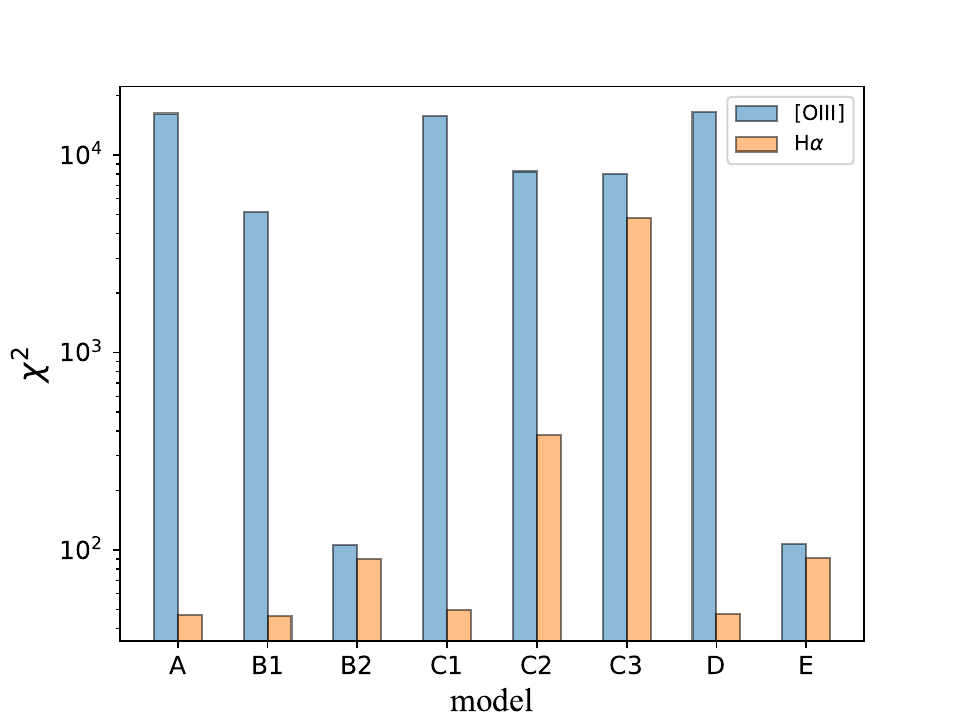}
    \includegraphics[width=0.49\linewidth]{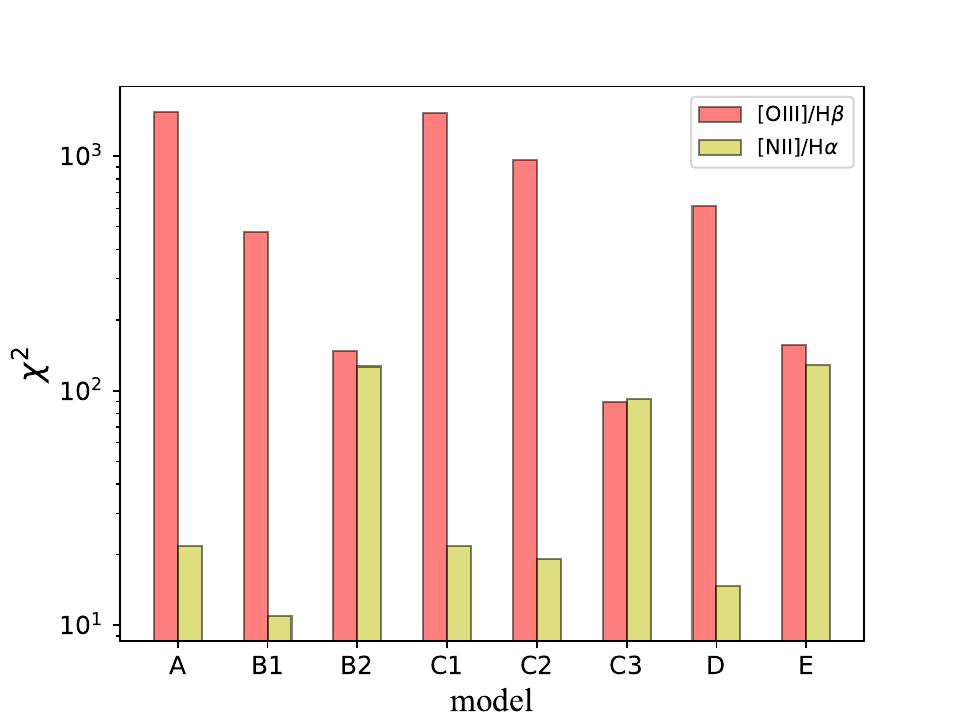}
    \caption{The $\chi^2$ distribution for models A-E in comparison with line intensities (left) and line ratios (right).}
    \label{fig:chi2}
\end{figure*}

We also performed a weighted least-square $\chi^2$ analysis to quantify the discrepancies between the models and the observations. We note that the $\chi^2$ values are only meaningful in a relative sense. As shown in Figure \ref{fig:chi2}, the model-predicted [OIII] flux exhibits substantially larger deviations from the data than H$\alpha$ in all cases except Model B2 and E, where the DIG-like component dominates the [OIII] emission. For the line ratios, the $\chi^{2}$ value of [O\,\textsc{iii}]/H$\beta$ is reduced from $\gtrsim 10^{3}$ in the fiducial Model A to $\sim 10^{2}$ in Models~B2, C3, and~E, whereas the $\chi^{2}$ of [N,\textsc{ii}]/H$\alpha$ increases from $\sim 10$ to $\sim 10^{2}$. This behavior highlights the intrinsic tension in reproducing both high- and low-ionization diagnostics simultaneously: models that improve the agreement for [O\,\textsc{iii}] generally do so at the expense of matching the low-ionization lines.




\bibliography{ref}{}

@ARTICLE{2018AJ....156..123A,
       author = {{Astropy Collaboration} and {Price-Whelan}, A.~M. and {Sip{\H{o}}cz}, B.~M. and {G{\"u}nther}, H.~M. and {Lim}, P.~L. and {Crawford}, S.~M. and {Conseil}, S. and {Shupe}, D.~L. and {Craig}, M.~W. and {Dencheva}, N. and {Ginsburg}, A. and {VanderPlas}, J.~T. and {Bradley}, L.~D. and {P{\'e}rez-Su{\'a}rez}, D. and {de Val-Borro}, M. and {Aldcroft}, T.~L. and {Cruz}, K.~L. and {Robitaille}, T.~P. and {Tollerud}, E.~J. and {Ardelean}, C. and {Babej}, T. and {Bach}, Y.~P. and {Bachetti}, M. and {Bakanov}, A.~V. and {Bamford}, S.~P. and {Barentsen}, G. and {Barmby}, P. and {Baumbach}, A. and {Berry}, K.~L. and {Biscani}, F. and {Boquien}, M. and {Bostroem}, K.~A. and {Bouma}, L.~G. and {Brammer}, G.~B. and {Bray}, E.~M. and {Breytenbach}, H. and {Buddelmeijer}, H. and {Burke}, D.~J. and {Calderone}, G. and {Cano Rodr{\'\i}guez}, J.~L. and {Cara}, M. and {Cardoso}, J.~V.~M. and {Cheedella}, S. and {Copin}, Y. and {Corrales}, L. and {Crichton}, D. and {D'Avella}, D. and {Deil}, C. and {Depagne}, {\'E}. and {Dietrich}, J.~P. and {Donath}, A. and {Droettboom}, M. and {Earl}, N. and {Erben}, T. and {Fabbro}, S. and {Ferreira}, L.~A. and {Finethy}, T. and {Fox}, R.~T. and {Garrison}, L.~H. and {Gibbons}, S.~L.~J. and {Goldstein}, D.~A. and {Gommers}, R. and {Greco}, J.~P. and {Greenfield}, P. and {Groener}, A.~M. and {Grollier}, F. and {Hagen}, A. and {Hirst}, P. and {Homeier}, D. and {Horton}, A.~J. and {Hosseinzadeh}, G. and {Hu}, L. and {Hunkeler}, J.~S. and {Ivezi{\'c}}, {\v{Z}}. and {Jain}, A. and {Jenness}, T. and {Kanarek}, G. and {Kendrew}, S. and {Kern}, N.~S. and {Kerzendorf}, W.~E. and {Khvalko}, A. and {King}, J. and {Kirkby}, D. and {Kulkarni}, A.~M. and {Kumar}, A. and {Lee}, A. and {Lenz}, D. and {Littlefair}, S.~P. and {Ma}, Z. and {Macleod}, D.~M. and {Mastropietro}, M. and {McCully}, C. and {Montagnac}, S. and {Morris}, B.~M. and {Mueller}, M. and {Mumford}, S.~J. and {Muna}, D. and {Murphy}, N.~A. and {Nelson}, S. and {Nguyen}, G.~H. and {Ninan}, J.~P. and {N{\"o}the}, M. and {Ogaz}, S. and {Oh}, S. and {Parejko}, J.~K. and {Parley}, N. and {Pascual}, S. and {Patil}, R. and {Patil}, A.~A. and {Plunkett}, A.~L. and {Prochaska}, J.~X. and {Rastogi}, T. and {Reddy Janga}, V. and {Sabater}, J. and {Sakurikar}, P. and {Seifert}, M. and {Sherbert}, L.~E. and {Sherwood-Taylor}, H. and {Shih}, A.~Y. and {Sick}, J. and {Silbiger}, M.~T. and {Singanamalla}, S. and {Singer}, L.~P. and {Sladen}, P.~H. and {Sooley}, K.~A. and {Sornarajah}, S. and {Streicher}, O. and {Teuben}, P. and {Thomas}, S.~W. and {Tremblay}, G.~R. and {Turner}, J.~E.~H. and {Terr{\'o}n}, V. and {van Kerkwijk}, M.~H. and {de la Vega}, A. and {Watkins}, L.~L. and {Weaver}, B.~A. and {Whitmore}, J.~B. and {Woillez}, J. and {Zabalza}, V. and {Astropy Contributors}},
        title = "{The Astropy Project: Building an Open-science Project and Status of the v2.0 Core Package}",
      journal = {\aj},
         year = 2018,
        month = sep,
       volume = {156},
       number = {3},
          eid = {123},
        pages = {123},
          doi = {10.3847/1538-3881/aabc4f},
archivePrefix = {arXiv},
       eprint = {1801.02634},
 primaryClass = {astro-ph.IM},
       adsurl = {https://ui.adsabs.harvard.edu/abs/2018AJ....156..123A}
}

@ARTICLE{2013A&A...558A..33A,
       author = {{Astropy Collaboration} and {Robitaille}, Thomas P. and {Tollerud}, Erik J. and {Greenfield}, Perry and {Droettboom}, Michael and {Bray}, Erik and {Aldcroft}, Tom and {Davis}, Matt and {Ginsburg}, Adam and {Price-Whelan}, Adrian M. and {Kerzendorf}, Wolfgang E. and {Conley}, Alexander and {Crighton}, Neil and {Barbary}, Kyle and {Muna}, Demitri and {Ferguson}, Henry and {Grollier}, Fr{\'e}d{\'e}ric and {Parikh}, Madhura M. and {Nair}, Prasanth H. and {Unther}, Hans M. and {Deil}, Christoph and {Woillez}, Julien and {Conseil}, Simon and {Kramer}, Roban and {Turner}, James E.~H. and {Singer}, Leo and {Fox}, Ryan and {Weaver}, Benjamin A. and {Zabalza}, Victor and {Edwards}, Zachary I. and {Azalee Bostroem}, K. and {Burke}, D.~J. and {Casey}, Andrew R. and {Crawford}, Steven M. and {Dencheva}, Nadia and {Ely}, Justin and {Jenness}, Tim and {Labrie}, Kathleen and {Lim}, Pey Lian and {Pierfederici}, Francesco and {Pontzen}, Andrew and {Ptak}, Andy and {Refsdal}, Brian and {Servillat}, Mathieu and {Streicher}, Ole},
        title = "{Astropy: A community Python package for astronomy}",
      journal = {\aap},
         year = 2013,
        month = oct,
       volume = {558},
          eid = {A33},
        pages = {A33},
          doi = {10.1051/0004-6361/201322068},
archivePrefix = {arXiv},
       eprint = {1307.6212},
 primaryClass = {astro-ph.IM},
       adsurl = {https://ui.adsabs.harvard.edu/abs/2013A&A...558A..33A}
}

@ARTICLE{2019MNRAS.484..964L,
       author = {{Li}, Zongnan and {Li}, Zhiyuan and {Zhou}, Ping and {Gao}, Yu and {Jiang}, Xue-Jian and {Dong}, Hui},
        title = "{JCMT mapping of CO(3-2) in the circumnuclear region of M31}",
      journal = {\mnras},
         year = 2019,
        month = mar,
       volume = {484},
       number = {1},
        pages = {964-973},
          doi = {10.1093/mnras/stz040},
archivePrefix = {arXiv},
       eprint = {1812.10887},
 primaryClass = {astro-ph.GA},
       adsurl = {https://ui.adsabs.harvard.edu/abs/2019MNRAS.484..964L}
}

@ARTICLE{2023ApJ...958...89L,
       author = {{Li}, Zongnan and {Li}, Zhiyuan and {Garc{\'\i}a-Benito}, Rub{\'e}n and {Jin}, Yifei},
        title = "{CAHA/PPAK Integral-field Spectroscopic Observations of M81. II. Testing Photoionization Models in a Spatially Resolved LINER}",
      journal = {\apj},
         year = 2023,
        month = nov,
       volume = {958},
       number = {1},
          eid = {89},
        pages = {89},
          doi = {10.3847/1538-4357/ad0299},
archivePrefix = {arXiv},
       eprint = {2310.07304},
 primaryClass = {astro-ph.GA},
       adsurl = {https://ui.adsabs.harvard.edu/abs/2023ApJ...958...89L}
}

@ARTICLE{2015ApJ...805..183L,
       author = {{Lewis}, Alexia R. and {Dolphin}, Andrew E. and {Dalcanton}, Julianne J. and {Weisz}, Daniel R. and {Williams}, Benjamin F. and {Bell}, Eric F. and {Seth}, Anil C. and {Simones}, Jacob E. and {Skillman}, Evan D. and {Choi}, Yumi and {Fouesneau}, Morgan and {Guhathakurta}, Puragra and {Johnson}, Lent C. and {Kalirai}, Jason S. and {Leroy}, Adam K. and {Monachesi}, Antonela and {Rix}, Hans-Walter and {Schruba}, Andreas},
        title = "{The Panchromatic Hubble Andromeda Treasury. XI. The Spatially Resolved Recent Star Formation History of M31}",
      journal = {\apj},
         year = 2015,
        month = jun,
       volume = {805},
       number = {2},
          eid = {183},
        pages = {183},
          doi = {10.1088/0004-637X/805/2/183},
archivePrefix = {arXiv},
       eprint = {1504.03338},
 primaryClass = {astro-ph.GA},
       adsurl = {https://ui.adsabs.harvard.edu/abs/2015ApJ...805..183L}
}

@ARTICLE{2018A&A...611A..38O,
       author = {{Opitsch}, M. and {Fabricius}, M.~H. and {Saglia}, R.~P. and {Bender}, R. and {Bla{\~n}a}, M. and {Gerhard}, O.},
        title = "{Evidence for non-axisymmetry in M 31 from wide-field kinematics of stars and gas}",
      journal = {\aap},
         year = 2018,
        month = mar,
       volume = {611},
          eid = {A38},
        pages = {A38},
          doi = {10.1051/0004-6361/201730597},
archivePrefix = {arXiv},
       eprint = {1707.06652},
 primaryClass = {astro-ph.GA},
       adsurl = {https://ui.adsabs.harvard.edu/abs/2018A&A...611A..38O}
}

@ARTICLE{2006A&A...460...45G,
       author = {{Gonz{\'a}lez-Mart{\'\i}n}, O. and {Masegosa}, J. and {M{\'a}rquez}, I. and {Guerrero}, M.~A. and {Dultzin-Hacyan}, D.},
        title = "{X-ray nature of the LINER nuclear sources}",
      journal = {\aap},
         year = 2006,
        month = dec,
       volume = {460},
       number = {1},
        pages = {45-57},
          doi = {10.1051/0004-6361:20054756},
archivePrefix = {arXiv},
       eprint = {astro-ph/0605629},
 primaryClass = {astro-ph},
       adsurl = {https://ui.adsabs.harvard.edu/abs/2006A&A...460...45G}
}

@ARTICLE{1997ApJ...487..568H,
       author = {{Ho}, Luis C. and {Filippenko}, Alexei V. and {Sargent}, Wallace L.~W.},
        title = "{A Search for ``Dwarf'' Seyfert Nuclei. V. Demographics of Nuclear Activity in Nearby Galaxies}",
      journal = {\apj},
         year = 1997,
        month = oct,
       volume = {487},
       number = {2},
        pages = {568-578},
          doi = {10.1086/304638},
archivePrefix = {arXiv},
       eprint = {astro-ph/9704108},
 primaryClass = {astro-ph},
       adsurl = {https://ui.adsabs.harvard.edu/abs/1997ApJ...487..568H}
}

@ARTICLE{1984ApJ...286...42F,
       author = {{Ferland}, G.~J. and {Mushotzky}, R.~F.},
        title = "{Osmic rays and the emission-line regions of active galactic nuclei.}",
      journal = {\apj},
         year = 1984,
        month = nov,
       volume = {286},
        pages = {42-52},
          doi = {10.1086/162574},
       adsurl = {https://ui.adsabs.harvard.edu/abs/1984ApJ...286...42F}
}

@ARTICLE{2023RMxAA..59..327C,
       author = {{Chatzikos}, M. and {Bianchi}, S. and {Camilloni}, F. and {Chakraborty}, P. and {Gunasekera}, C.~M. and {Guzm{\'a}n}, F. and {Milby}, J.~S. and {Sarkar}, A. and {Shaw}, G. and {van Hoof}, P.~A.~M. and {Ferland}, G.~J.},
        title = "{The 2023 Release of Cloudy}",
      journal = {\rmxaa},
         year = 2023,
        month = oct,
       volume = {59},
        pages = {327-343},
          doi = {10.22201/ia.01851101p.2023.59.02.12},
archivePrefix = {arXiv},
       eprint = {2308.06396},
 primaryClass = {astro-ph.GA},
       adsurl = {https://ui.adsabs.harvard.edu/abs/2023RMxAA..59..327C}
}

@ARTICLE{2011ApJ...739...20C,
       author = {{Courteau}, St{\'e}phane and {Widrow}, Lawrence M. and {McDonald}, Michael and {Guhathakurta}, Puragra and {Gilbert}, Karoline M. and {Zhu}, Yucong and {Beaton}, Rachael Lynn and {Majewski}, Steven R.},
        title = "{The Luminosity Profile and Structural Parameters of the Andromeda Galaxy}",
      journal = {\apj},
         year = 2011,
        month = sep,
       volume = {739},
       number = {1},
          eid = {20},
        pages = {20},
          doi = {10.1088/0004-637X/739/1/20},
archivePrefix = {arXiv},
       eprint = {1106.3564},
 primaryClass = {astro-ph.CO},
       adsurl = {https://ui.adsabs.harvard.edu/abs/2011ApJ...739...20C}
}

@ARTICLE{2018MNRAS.481.3210B,
       author = {{Bla{\~n}a D{\'\i}az}, Mat{\'\i}as and {Gerhard}, Ortwin and {Wegg}, Christopher and {Portail}, Matthieu and {Opitsch}, Michael and {Saglia}, Roberto and {Fabricius}, Maximilian and {Erwin}, Peter and {Bender}, Ralf},
        title = "{Sculpting Andromeda - made-to-measure models for M31's bar and composite bulge: dynamics, stellar and dark matter mass}",
      journal = {\mnras},
         year = 2018,
        month = dec,
       volume = {481},
       number = {3},
        pages = {3210-3243},
          doi = {10.1093/mnras/sty2311},
archivePrefix = {arXiv},
       eprint = {1808.07494},
 primaryClass = {astro-ph.GA},
       adsurl = {https://ui.adsabs.harvard.edu/abs/2018MNRAS.481.3210B}
}

@BOOK{1989agna.book.....O,
       author = {{Osterbrock}, Donald E.},
        title = "{Astrophysics of gaseous nebulae and active galactic nuclei}",
         year = 1989,
       adsurl = {https://ui.adsabs.harvard.edu/abs/1989agna.book.....O}
}

@ARTICLE{2009MNRAS.399.1026B,
       author = {{Buckle}, J.~V. and {Hills}, R.~E. and {Smith}, H. and {Dent}, W.~R.~F. and {Bell}, G. and {Curtis}, E.~I. and {Dace}, R. and {Gibson}, H. and {Graves}, S.~F. and {Leech}, J. and {Richer}, J.~S. and {Williamson}, R. and {Withington}, S. and {Yassin}, G. and {Bennett}, R. and {Hastings}, P. and {Laidlaw}, I. and {Lightfoot}, J.~F. and {Burgess}, T. and {Dewdney}, P.~E. and {Hovey}, G. and {Willis}, A.~G. and {Redman}, R. and {Wooff}, B. and {Berry}, D.~S. and {Cavanagh}, B. and {Davis}, G.~R. and {Dempsey}, J. and {Friberg}, P. and {Jenness}, T. and {Kackley}, R. and {Rees}, N.~P. and {Tilanus}, R. and {Walther}, C. and {Zwart}, W. and {Klapwijk}, T.~M. and {Kroug}, M. and {Zijlstra}, T.},
        title = "{HARP/ACSIS: a submillimetre spectral imaging system on the James Clerk Maxwell Telescope}",
      journal = {\mnras},
         year = 2009,
        month = oct,
       volume = {399},
       number = {2},
        pages = {1026-1043},
          doi = {10.1111/j.1365-2966.2009.15347.x},
archivePrefix = {arXiv},
       eprint = {0907.3610},
 primaryClass = {astro-ph.IM},
       adsurl = {https://ui.adsabs.harvard.edu/abs/2009MNRAS.399.1026B}
}

@ARTICLE{2009MNRAS.397..148L,
       author = {{Li}, Zhiyuan and {Wang}, Q. Daniel and {Wakker}, Bart P.},
        title = "{M31* and its circumnuclear environment}",
      journal = {\mnras},
         year = 2009,
        month = jul,
       volume = {397},
       number = {1},
        pages = {148-163},
          doi = {10.1111/j.1365-2966.2009.14918.x},
archivePrefix = {arXiv},
       eprint = {0902.3847},
 primaryClass = {astro-ph.GA},
       adsurl = {https://ui.adsabs.harvard.edu/abs/2009MNRAS.397..148L}
}

@BOOK{1968adga.book.....S,
       author = {{Sersic}, Jose Luis},
        title = "{Atlas de Galaxias Australes}",
         year = 1968,
       adsurl = {https://ui.adsabs.harvard.edu/abs/1968adga.book.....S}
}

@ARTICLE{2016MNRAS.459.2262D,
       author = {{Dong}, Hui and {Li}, Zhiyuan and {Wang}, Q.~D. and {Lauer}, Tod R. and {Olsen}, Knut A.~G. and {Saha}, Abhijit and {Dalcanton}, Julianne J. and {Groves}, Brent A.},
        title = "{High-resolution mapping of dust via extinction in the M31 bulge}",
      journal = {\mnras},
         year = 2016,
        month = jun,
       volume = {459},
       number = {2},
        pages = {2262-2273},
          doi = {10.1093/mnras/stw778},
archivePrefix = {arXiv},
       eprint = {1603.09637},
 primaryClass = {astro-ph.GA},
       adsurl = {https://ui.adsabs.harvard.edu/abs/2016MNRAS.459.2262D}
}

@article{Baldwin_1981,
	doi = {10.1086/130766},
	url = {https://doi.org/10.1086/130766},
	year = 1981,
	month = {feb},
	publisher = {{IOP} Publishing},
	volume = {93},
	pages = {5},
	author = {J. A. Baldwin and M. M. Phillips and R. Terlevich},
	title = {Classification parameters for the emission-line spectra of extragalactic objects},
	journal = {Publications of the Astronomical Society of the Pacific}
}

@ARTICLE{2009ApJ...705.1395C,
       author = {{Chemin}, Laurent and {Carignan}, Claude and {Foster}, Tyler},
        title = "{H I Kinematics and Dynamics of Messier 31}",
      journal = {\apj},
         year = 2009,
        month = nov,
       volume = {705},
       number = {2},
        pages = {1395-1415},
          doi = {10.1088/0004-637X/705/2/1395},
archivePrefix = {arXiv},
       eprint = {0909.3846},
 primaryClass = {astro-ph.CO},
       adsurl = {https://ui.adsabs.harvard.edu/abs/2009ApJ...705.1395C}
}

@ARTICLE{2015MNRAS.451.4126D,
       author = {{Dong}, Hui and {Li}, Zhiyuan and {Wang}, Q. Daniel and {Lauer}, Tod R. and {Olsen}, Knut A.~G. and {Saha}, Abhijit and {Dalcanton}, Julianne J. and {Williams}, Benjamin F.},
        title = "{Photometric evidence of an intermediate-age stellar population in the inner bulge of M31}",
      journal = {\mnras},
         year = 2015,
        month = aug,
       volume = {451},
       number = {4},
        pages = {4126-4138},
          doi = {10.1093/mnras/stv1256},
archivePrefix = {arXiv},
       eprint = {1506.01097},
 primaryClass = {astro-ph.GA},
       adsurl = {https://ui.adsabs.harvard.edu/abs/2015MNRAS.451.4126D}
}

@ARTICLE{2011MNRAS.413.1687C,
       author = {{Cid Fernandes}, R. and {Stasi{\'n}ska}, G. and {Mateus}, A. and {Vale Asari}, N.},
        title = "{A comprehensive classification of galaxies in the Sloan Digital Sky Survey: how to tell true from fake AGN?}",
      journal = {\mnras},
         year = 2011,
        month = may,
       volume = {413},
       number = {3},
        pages = {1687-1699},
          doi = {10.1111/j.1365-2966.2011.18244.x},
archivePrefix = {arXiv},
       eprint = {1012.4426},
 primaryClass = {astro-ph.CO},
       adsurl = {https://ui.adsabs.harvard.edu/abs/2011MNRAS.413.1687C}
}

@ARTICLE{2014A&A...561A..10P,
       author = {{Proxauf}, B. and {{\"O}ttl}, S. and {Kimeswenger}, S.},
        title = "{Upgrading electron temperature and electron density diagnostic diagrams of forbidden line emission}",
      journal = {\aap},
         year = 2014,
        month = jan,
       volume = {561},
          eid = {A10},
        pages = {A10},
          doi = {10.1051/0004-6361/201322581},
archivePrefix = {arXiv},
       eprint = {1311.5041},
 primaryClass = {astro-ph.IM},
       adsurl = {https://ui.adsabs.harvard.edu/abs/2014A&A...561A..10P}
}

@ARTICLE{2015ApJS..219....5Q,
       author = {{Querejeta}, Miguel and {Meidt}, Sharon E. and {Schinnerer}, Eva and {Cisternas}, Mauricio and {Mu{\~n}oz-Mateos}, Juan Carlos and {Sheth}, Kartik and {Knapen}, Johan and {van de Ven}, Glenn and {Norris}, Mark A. and {Peletier}, Reynier and {Laurikainen}, Eija and {Salo}, Heikki and {Holwerda}, Benne W. and {Athanassoula}, E. and {Bosma}, Albert and {Groves}, Brent and {Ho}, Luis C. and {Gadotti}, Dimitri A. and {Zaritsky}, Dennis and {Regan}, Michael and {Hinz}, Joannah and {Gil de Paz}, Armando and {Menendez-Delmestre}, Karin and {Seibert}, Mark and {Mizusawa}, Trisha and {Kim}, Taehyun and {Erroz-Ferrer}, Santiago and {Laine}, Jarkko and {Comer{\'o}n}, S{\'e}bastien},
        title = "{The Spitzer Survey of Stellar Structure in Galaxies (S$^{4}$G): Precise Stellar Mass Distributions from Automated Dust Correction at 3.6 {\ensuremath{\mu}}m}",
      journal = {\apjs},
         year = 2015,
        month = jul,
       volume = {219},
       number = {1},
          eid = {5},
        pages = {5},
          doi = {10.1088/0067-0049/219/1/5},
archivePrefix = {arXiv},
       eprint = {1410.0009},
 primaryClass = {astro-ph.GA},
       adsurl = {https://ui.adsabs.harvard.edu/abs/2015ApJS..219....5Q}
}

@ARTICLE{2016MNRAS.461.3111B,
       author = {{Belfiore}, Francesco and {Maiolino}, Roberto and {Maraston}, Claudia and {Emsellem}, Eric and {Bershady}, Matthew A. and {Masters}, Karen L. and {Yan}, Renbin and {Bizyaev}, Dmitry and {Boquien}, M{\'e}d{\'e}ric and {Brownstein}, Joel R. and {Bundy}, Kevin and {Drory}, Niv and {Heckman}, Timothy M. and {Law}, David R. and {Roman-Lopes}, Alexandre and {Pan}, Kaike and {Stanghellini}, Letizia and {Thomas}, Daniel and {Weijmans}, Anne-Marie and {Westfall}, Kyle B.},
        title = "{SDSS IV MaNGA - spatially resolved diagnostic diagrams: a proof that many galaxies are LIERs}",
      journal = {\mnras},
         year = 2016,
        month = sep,
       volume = {461},
       number = {3},
        pages = {3111-3134},
          doi = {10.1093/mnras/stw1234},
archivePrefix = {arXiv},
       eprint = {1605.07189},
 primaryClass = {astro-ph.GA},
       adsurl = {https://ui.adsabs.harvard.edu/abs/2016MNRAS.461.3111B}
}

@ARTICLE{2013A&A...558A..43S,
       author = {{Singh}, R. and {van de Ven}, G. and {Jahnke}, K. and {Lyubenova}, M. and {Falc{\'o}n-Barroso}, J. and {Alves}, J. and {Cid Fernandes}, R. and {Galbany}, L. and {Garc{\'\i}a-Benito}, R. and {Husemann}, B. and {Kennicutt}, R.~C. and {Marino}, R.~A. and {M{\'a}rquez}, I. and {Masegosa}, J. and {Mast}, D. and {Pasquali}, A. and {S{\'a}nchez}, S.~F. and {Walcher}, J. and {Wild}, V. and {Wisotzki}, L. and {Ziegler}, B.},
        title = "{The nature of LINER galaxies:. Ubiquitous hot old stars and rare accreting black holes}",
      journal = {\aap},
         year = 2013,
        month = oct,
       volume = {558},
          eid = {A43},
        pages = {A43},
          doi = {10.1051/0004-6361/201322062},
archivePrefix = {arXiv},
       eprint = {1308.4271},
 primaryClass = {astro-ph.GA},
       adsurl = {https://ui.adsabs.harvard.edu/abs/2013A&A...558A..43S}
}

@ARTICLE{2025A&A...695A.194C,
       author = {{Cros}, Lucie and {Combes}, Fran{\c{c}}oise and {Melchior}, Anne-Laure and {Martin}, Thomas},
        title = "{Central kiloparsec region of Andromeda: I. Dynamical modeling}",
      journal = {\aap},
         year = 2025,
        month = mar,
       volume = {695},
          eid = {A194},
        pages = {A194},
          doi = {10.1051/0004-6361/202453067},
archivePrefix = {arXiv},
       eprint = {2411.18460},
 primaryClass = {astro-ph.GA},
       adsurl = {https://ui.adsabs.harvard.edu/abs/2025A&A...695A.194C}
}

@ARTICLE{2011MNRAS.415.2182F,
       author = {{Flores-Fajardo}, N. and {Morisset}, C. and {Stasi{\'n}ska}, G. and {Binette}, L.},
        title = "{Ionization of the diffuse gas in galaxies: hot low-mass evolved stars at work}",
      journal = {\mnras},
         year = 2011,
        month = aug,
       volume = {415},
       number = {3},
        pages = {2182-2192},
          doi = {10.1111/j.1365-2966.2011.18848.x},
archivePrefix = {arXiv},
       eprint = {1104.0525},
 primaryClass = {astro-ph.GA},
       adsurl = {https://ui.adsabs.harvard.edu/abs/2011MNRAS.415.2182F}
}

@ARTICLE{1985ApJ...290..136J,
       author = {{Jacoby}, G.~H. and {Ford}, H. and {Ciardullo}, R.},
        title = "{Ionized gas in the center of M 31.}",
      journal = {\apj},
         year = 1985,
        month = mar,
       volume = {290},
        pages = {136-139},
          doi = {10.1086/162968},
       adsurl = {https://ui.adsabs.harvard.edu/abs/1985ApJ...290..136J}
}

@ARTICLE{2018MNRAS.473.4130M,
       author = {{Martin}, Thomas B. and {Drissen}, Laurent and {Melchior}, Anne-Laure},
        title = "{A SITELLE view of M31's central region - I. Calibrations and radial velocity catalogue of nearly 800 emission-line point-like sources}",
      journal = {\mnras},
         year = 2018,
        month = jan,
       volume = {473},
       number = {3},
        pages = {4130-4149},
          doi = {10.1093/mnras/stx2513},
archivePrefix = {arXiv},
       eprint = {1707.01366},
 primaryClass = {astro-ph.GA},
       adsurl = {https://ui.adsabs.harvard.edu/abs/2018MNRAS.473.4130M}
}

@INPROCEEDINGS{2015ASPC..495..327M,
       author = {{Martin}, T. and {Drissen}, L. and {Joncas}, G.},
        title = "{ORBS, ORCS, OACS, a Software Suite for Data Reduction and Analysis of the Hyperspectral Imagers SITELLE and SpIOMM}",
    booktitle = {Astronomical Data Analysis Software an Systems XXIV (ADASS XXIV)},
         year = 2015,
       editor = {{Taylor}, A.~R. and {Rosolowsky}, E.},
       series = {Astronomical Society of the Pacific Conference Series},
       volume = {495},
        month = sep,
        pages = {327},
       adsurl = {https://ui.adsabs.harvard.edu/abs/2015ASPC..495..327M}
}

@INPROCEEDINGS{2016sf2a.conf...23M,
       author = {{Martin}, T.~B. and {Drissen}, L.},
        title = "{SITELLE's Data Release 1}",
    booktitle = {SF2A-2016: Proceedings of the Annual meeting of the French Society of Astronomy and Astrophysics},
         year = 2016,
       editor = {{Reyl{\'e}}, C. and {Richard}, J. and {Cambr{\'e}sy}, L. and {Deleuil}, M. and {P{\'e}contal}, E. and {Tresse}, L. and {Vauglin}, I.},
        month = dec,
        pages = {23-28},
       adsurl = {https://ui.adsabs.harvard.edu/abs/2016sf2a.conf...23M}
}

@INPROCEEDINGS{2012SPIE.8451E..3KM,
       author = {{Martin}, T. and {Drissen}, L. and {Joncas}, G.},
        title = "{ORBS: A data reduction software for the imaging Fourier transform spectrometers SpIOMM and SITELLE}",
    booktitle = {Software and Cyberinfrastructure for Astronomy II},
         year = 2012,
       editor = {{Radziwill}, Nicole M. and {Chiozzi}, Gianluca},
       series = {Society of Photo-Optical Instrumentation Engineers (SPIE) Conference Series},
       volume = {8451},
        month = sep,
          eid = {84513K},
        pages = {84513K},
          doi = {10.1117/12.925420},
       adsurl = {https://ui.adsabs.harvard.edu/abs/2012SPIE.8451E..3KM}
}

@ARTICLE{2016MNRAS.463.3409V,
       author = {{Vazdekis}, A. and {Koleva}, M. and {Ricciardelli}, E. and {R{\"o}ck}, B. and {Falc{\'o}n-Barroso}, J.},
        title = "{UV-extended E-MILES stellar population models: young components in massive early-type galaxies}",
      journal = {\mnras},
         year = 2016,
        month = dec,
       volume = {463},
       number = {4},
        pages = {3409-3436},
          doi = {10.1093/mnras/stw2231},
archivePrefix = {arXiv},
       eprint = {1612.01187},
 primaryClass = {astro-ph.GA},
       adsurl = {https://ui.adsabs.harvard.edu/abs/2016MNRAS.463.3409V}
}

@ARTICLE{2004PASP..116..138C,
       author = {{Cappellari}, Michele and {Emsellem}, Eric},
        title = "{Parametric Recovery of Line-of-Sight Velocity Distributions from Absorption-Line Spectra of Galaxies via Penalized Likelihood}",
      journal = {\pasp},
         year = 2004,
        month = feb,
       volume = {116},
       number = {816},
        pages = {138-147},
          doi = {10.1086/381875},
archivePrefix = {arXiv},
       eprint = {astro-ph/0312201},
 primaryClass = {astro-ph},
       adsurl = {https://ui.adsabs.harvard.edu/abs/2004PASP..116..138C}
}

@ARTICLE{2013A&A...549A..27M,
       author = {{Melchior}, A. -L. and {Combes}, F.},
        title = "{A cold-gas reservoir to fuel the M 31 nuclear black hole and stellar cluster}",
      journal = {\aap},
         year = 2013,
        month = jan,
       volume = {549},
          eid = {A27},
        pages = {A27},
          doi = {10.1051/0004-6361/201220204},
archivePrefix = {arXiv},
       eprint = {1210.4316},
 primaryClass = {astro-ph.CO},
       adsurl = {https://ui.adsabs.harvard.edu/abs/2013A&A...549A..27M}
}

@ARTICLE{2019MNRAS.485.3930D,
       author = {{Drissen}, Laurent and {Martin}, Thomas and {Rousseau-Nepton}, Laurie and {Robert}, Carmelle and {Martin}, R. Pierre and {Baril}, Marc and {Prunet}, Simon and {Joncas}, Gilles and {Thibault}, Simon and {Brousseau}, Denis and {Mandar}, Julie and {Grandmont}, Fr{\'e}d{\'e}ric and {Yee}, Howard and {Simard}, Luc},
        title = "{SITELLE: an Imaging Fourier Transform Spectrometer for the Canada-France-Hawaii Telescope}",
      journal = {\mnras},
         year = 2019,
        month = may,
       volume = {485},
       number = {3},
        pages = {3930-3946},
          doi = {10.1093/mnras/stz627},
archivePrefix = {arXiv},
       eprint = {1811.06644},
 primaryClass = {astro-ph.IM},
       adsurl = {https://ui.adsabs.harvard.edu/abs/2019MNRAS.485.3930D}
}

@ARTICLE{2009ApJ...699..486C,
       author = {{Conroy}, Charlie and {Gunn}, James E. and {White}, Martin},
        title = "{The Propagation of Uncertainties in Stellar Population Synthesis Modeling. I. The Relevance of Uncertain Aspects of Stellar Evolution and the Initial Mass Function to the Derived Physical Properties of Galaxies}",
      journal = {\apj},
         year = 2009,
        month = jul,
       volume = {699},
       number = {1},
        pages = {486-506},
          doi = {10.1088/0004-637X/699/1/486},
archivePrefix = {arXiv},
       eprint = {0809.4261},
 primaryClass = {astro-ph},
       adsurl = {https://ui.adsabs.harvard.edu/abs/2009ApJ...699..486C}
}

@ARTICLE{2010ApJ...712..833C,
       author = {{Conroy}, Charlie and {Gunn}, James E.},
        title = "{The Propagation of Uncertainties in Stellar Population Synthesis Modeling. III. Model Calibration, Comparison, and Evaluation}",
      journal = {\apj},
         year = 2010,
        month = apr,
       volume = {712},
       number = {2},
        pages = {833-857},
          doi = {10.1088/0004-637X/712/2/833},
archivePrefix = {arXiv},
       eprint = {0911.3151},
 primaryClass = {astro-ph.CO},
       adsurl = {https://ui.adsabs.harvard.edu/abs/2010ApJ...712..833C}
}

@ARTICLE{2012ApJ...755..131R,
       author = {{Rosenfield}, Philip and {Johnson}, L. Clifton and {Girardi}, L{\'e}o and {Dalcanton}, Julianne J. and {Bressan}, Alessandro and {Lang}, Dustin and {Williams}, Benjamin F. and {Guhathakurta}, Puragra and {Howley}, Kirsten M. and {Lauer}, Tod R. and {Bell}, Eric F. and {Bianchi}, Luciana and {Caldwell}, Nelson and {Dolphin}, Andrew and {Dorman}, Claire E. and {Gilbert}, Karoline M. and {Kalirai}, Jason and {Larsen}, S{\o}ren S. and {Olsen}, Knut A.~G. and {Rix}, Hans-Walter and {Seth}, Anil C. and {Skillman}, Evan D. and {Weisz}, Daniel R.},
        title = "{The Panchromatic Hubble Andromeda Treasury. I. Bright UV Stars in the Bulge of M31}",
      journal = {\apj},
         year = 2012,
        month = aug,
       volume = {755},
       number = {2},
          eid = {131},
        pages = {131},
          doi = {10.1088/0004-637X/755/2/131},
archivePrefix = {arXiv},
       eprint = {1206.4045},
 primaryClass = {astro-ph.CO},
       adsurl = {https://ui.adsabs.harvard.edu/abs/2012ApJ...755..131R}
}

@ARTICLE{2016ApJ...826..175H,
       author = {{Herrera-Camus}, R. and {Bolatto}, A. and {Smith}, J.~D. and {Draine}, B. and {Pellegrini}, E. and {Wolfire}, M. and {Croxall}, K. and {de Looze}, I. and {Calzetti}, D. and {Kennicutt}, R. and {Crocker}, A. and {Armus}, L. and {van der Werf}, P. and {Sandstrom}, K. and {Galametz}, M. and {Brandl}, B. and {Groves}, B. and {Rigopoulou}, D. and {Walter}, F. and {Leroy}, A. and {Boquien}, M. and {Tabatabaei}, F.~S. and {Beirao}, P.},
        title = "{The Ionized Gas in Nearby Galaxies as Traced by the [N II] 122 and 205 {\ensuremath{\mu}}m Transitions}",
      journal = {\apj},
         year = 2016,
        month = aug,
       volume = {826},
       number = {2},
          eid = {175},
        pages = {175},
          doi = {10.3847/0004-637X/826/2/175},
archivePrefix = {arXiv},
       eprint = {1605.03180},
 primaryClass = {astro-ph.GA},
       adsurl = {https://ui.adsabs.harvard.edu/abs/2016ApJ...826..175H}
}

@ARTICLE{2025arXiv250513677B,
       author = {{Burman}, Shivam and {Malik}, Sunil and {Singh}, Suprit and {Wadadekar}, Yogesh},
        title = "{Unveiling Electron Density Profile in Nearby Galaxies using SDSS MaNGA}",
      journal = {arXiv e-prints},
         year = 2025,
        month = may,
          eid = {arXiv:2505.13677},
        pages = {arXiv:2505.13677},
          doi = {10.48550/arXiv.2505.13677},
archivePrefix = {arXiv},
       eprint = {2505.13677},
 primaryClass = {astro-ph.GA},
       adsurl = {https://ui.adsabs.harvard.edu/abs/2025arXiv250513677B}
}

@ARTICLE{1997ApJ...483..666G,
       author = {{Greenawalt}, B. and {Walterbos}, R.~A.~M. and {Braun}, R.},
        title = "{Optical Spectroscopy of Diffuse Ionized Gas in M31}",
      journal = {\apj},
         year = 1997,
        month = jul,
       volume = {483},
       number = {2},
        pages = {666-674},
          doi = {10.1086/304285},
archivePrefix = {arXiv},
       eprint = {astro-ph/9612167},
 primaryClass = {astro-ph},
       adsurl = {https://ui.adsabs.harvard.edu/abs/1997ApJ...483..666G}
}

@ARTICLE{2017ApJ...845..155B,
       author = {{Boettcher}, Erin and {Gallagher}, III, J.~S. and {Zweibel}, Ellen G.},
        title = "{Detection of Extraplanar Diffuse Ionized Gas in M83}",
      journal = {\apj},
         year = 2017,
        month = aug,
       volume = {845},
       number = {2},
          eid = {155},
        pages = {155},
          doi = {10.3847/1538-4357/aa81ca},
archivePrefix = {arXiv},
       eprint = {1707.08126},
 primaryClass = {astro-ph.GA},
       adsurl = {https://ui.adsabs.harvard.edu/abs/2017ApJ...845..155B}
}

@ARTICLE{2014MNRAS.438.2804N,
       author = {{Nemmen}, Rodrigo S. and {Storchi-Bergmann}, Thaisa and {Eracleous}, Michael},
        title = "{Spectral models for low-luminosity active galactic nuclei in LINERs: the role of advection-dominated accretion and jets}",
      journal = {\mnras},
         year = 2014,
        month = mar,
       volume = {438},
       number = {4},
        pages = {2804-2827},
          doi = {10.1093/mnras/stt2388},
archivePrefix = {arXiv},
       eprint = {1312.1982},
 primaryClass = {astro-ph.HE},
       adsurl = {https://ui.adsabs.harvard.edu/abs/2014MNRAS.438.2804N}
}

@ARTICLE{2017ApJ...839...87B,
       author = {{Bizyaev}, D. and {Walterbos}, R.~A.~M. and {Yoachim}, P. and {Riffel}, R.~A. and {Fern{\'a}ndez-Trincado}, J.~G. and {Pan}, K. and {Diamond-Stanic}, A.~M. and {Jones}, A. and {Thomas}, D. and {Cleary}, J. and {Brinkmann}, J.},
        title = "{SDSS IV MaNGA{\textemdash}Rotation Velocity Lags in the Extraplanar Ionized Gas from MaNGA Observations of Edge-on Galaxies}",
      journal = {\apj},
         year = 2017,
        month = apr,
       volume = {839},
       number = {2},
          eid = {87},
        pages = {87},
          doi = {10.3847/1538-4357/aa6979},
archivePrefix = {arXiv},
       eprint = {1704.02582},
 primaryClass = {astro-ph.GA},
       adsurl = {https://ui.adsabs.harvard.edu/abs/2017ApJ...839...87B}
}

@ARTICLE{2022A&A...659A..26B,
       author = {{Belfiore}, F. and {Santoro}, F. and {Groves}, B. and {Schinnerer}, E. and {Kreckel}, K. and {Glover}, S.~C.~O. and {Klessen}, R.~S. and {Emsellem}, E. and {Blanc}, G.~A. and {Congiu}, E. and {Barnes}, A.~T. and {Boquien}, M. and {Chevance}, M. and {Dale}, D.~A. and {Kruijssen}, J.~M. Diederik and {Leroy}, A.~K. and {Pan}, H.-A. and {Pessa}, I. and {Schruba}, A. and {Williams}, T.~G.},
        title = "{A tale of two DIGs: The relative role of H II regions and low-mass hot evolved stars in powering the diffuse ionised gas (DIG) in PHANGS-MUSE galaxies}",
      journal = {\aap},
         year = 2022,
        month = mar,
       volume = {659},
          eid = {A26},
        pages = {A26},
          doi = {10.1051/0004-6361/202141859},
archivePrefix = {arXiv},
       eprint = {2111.14876},
 primaryClass = {astro-ph.GA},
       adsurl = {https://ui.adsabs.harvard.edu/abs/2022A&A...659A..26B}
}

@ARTICLE{2017A&A...599A.141J,
       author = {{Jones}, A. and {Kauffmann}, G. and {D'Souza}, R. and {Bizyaev}, D. and {Law}, D. and {Haffner}, L. and {Bah{\'e}}, Y. and {Andrews}, B. and {Bershady}, M. and {Brownstein}, J. and {Bundy}, K. and {Cherinka}, B. and {Diamond-Stanic}, A. and {Drory}, N. and {Riffel}, R.~A. and {S{\'a}nchez}, S.~F. and {Thomas}, D. and {Wake}, D. and {Yan}, R. and {Zhang}, K.},
        title = "{SDSS IV MaNGA: Deep observations of extra-planar, diffuse ionized gas around late-type galaxies from stacked IFU spectra}",
      journal = {\aap},
         year = 2017,
        month = mar,
       volume = {599},
          eid = {A141},
        pages = {A141},
          doi = {10.1051/0004-6361/201629802},
archivePrefix = {arXiv},
       eprint = {1612.03920},
 primaryClass = {astro-ph.GA},
       adsurl = {https://ui.adsabs.harvard.edu/abs/2017A&A...599A.141J}
}

@ARTICLE{2021MNRAS.504.3013L,
       author = {{Li}, Anqi and {Marasco}, Antonino and {Fraternali}, Filippo and {Trager}, Scott and {Verheijen}, Marc A.~W.},
        title = "{A kinematic analysis of ionized extraplanar gas in the spiral galaxies NGC 3982 and NGC 4152}",
      journal = {\mnras},
         year = 2021,
        month = jun,
       volume = {504},
       number = {2},
        pages = {3013-3028},
          doi = {10.1093/mnras/stab1043},
archivePrefix = {arXiv},
       eprint = {2104.05736},
 primaryClass = {astro-ph.GA},
       adsurl = {https://ui.adsabs.harvard.edu/abs/2021MNRAS.504.3013L}
}

@ARTICLE{2019MNRAS.487..737J,
       author = {{Ji}, Suoqing and {Oh}, S. Peng and {Masterson}, Phillip},
        title = "{Simulations of radiative turbulent mixing layers}",
      journal = {\mnras},
         year = 2019,
        month = jul,
       volume = {487},
       number = {1},
        pages = {737-754},
          doi = {10.1093/mnras/stz1248},
archivePrefix = {arXiv},
       eprint = {1809.09101},
 primaryClass = {astro-ph.GA},
       adsurl = {https://ui.adsabs.harvard.edu/abs/2019MNRAS.487..737J}
}

@ARTICLE{2009ApJ...699..626H,
       author = {{Ho}, Luis C.},
        title = "{Radiatively Inefficient Accretion in Nearby Galaxies}",
      journal = {\apj},
         year = 2009,
        month = jul,
       volume = {699},
       number = {1},
        pages = {626-637},
          doi = {10.1088/0004-637X/699/1/626},
archivePrefix = {arXiv},
       eprint = {0906.4104},
 primaryClass = {astro-ph.GA},
       adsurl = {https://ui.adsabs.harvard.edu/abs/2009ApJ...699..626H}
}

@ARTICLE{2020MNRAS.491.4089D,
       author = {{den Brok}, Mark and {Carollo}, C. Marcella and {Erroz-Ferrer}, Santiago and {Fagioli}, Martina and {Brinchmann}, Jarle and {Emsellem}, Eric and {Krajnovi{\'c}}, Davor and {Marino}, Raffaella A. and {Onodera}, Masato and {Tacchella}, Sandro and {Weilbacher}, Peter M. and {Woo}, Joanna},
        title = "{The MUSE Atlas of Discs (MAD): Ionized gas kinematic maps and an application to diffuse ionized gas}",
      journal = {\mnras},
         year = 2020,
        month = jan,
       volume = {491},
       number = {3},
        pages = {4089-4107},
          doi = {10.1093/mnras/stz3184},
archivePrefix = {arXiv},
       eprint = {1911.06070},
 primaryClass = {astro-ph.GA},
       adsurl = {https://ui.adsabs.harvard.edu/abs/2020MNRAS.491.4089D}
}

@ARTICLE{2019ApJ...885..160B,
       author = {{Boettcher}, Erin and {Gallagher}, III, J.~S. and {Zweibel}, Ellen G.},
        title = "{A Dynamical Study of Extraplanar Diffuse Ionized Gas in NGC 5775}",
      journal = {\apj},
         year = 2019,
        month = nov,
       volume = {885},
       number = {2},
          eid = {160},
        pages = {160},
          doi = {10.3847/1538-4357/ab4904},
archivePrefix = {arXiv},
       eprint = {1909.11679},
 primaryClass = {astro-ph.GA},
       adsurl = {https://ui.adsabs.harvard.edu/abs/2019ApJ...885..160B}
}

@ARTICLE{2020ApJ...897..143K,
       author = {{Kado-Fong}, Erin and {Kim}, Jeong-Gyu and {Ostriker}, Eve C. and {Kim}, Chang-Goo},
        title = "{Diffuse Ionized Gas in Simulations of Multiphase, Star-forming Galactic Disks}",
      journal = {\apj},
         year = 2020,
        month = jul,
       volume = {897},
       number = {2},
          eid = {143},
        pages = {143},
          doi = {10.3847/1538-4357/ab9abd},
archivePrefix = {arXiv},
       eprint = {2006.06697},
 primaryClass = {astro-ph.GA},
       adsurl = {https://ui.adsabs.harvard.edu/abs/2020ApJ...897..143K}
}

@ARTICLE{2012MNRAS.427.1463Z,
       author = {{Zurita}, A. and {Bresolin}, F.},
        title = "{The chemical abundance in M31 from H II regions}",
      journal = {\mnras},
         year = 2012,
        month = dec,
       volume = {427},
       number = {2},
        pages = {1463-1481},
          doi = {10.1111/j.1365-2966.2012.22075.x},
archivePrefix = {arXiv},
       eprint = {1209.1505},
 primaryClass = {astro-ph.CO},
       adsurl = {https://ui.adsabs.harvard.edu/abs/2012MNRAS.427.1463Z}
}

@ARTICLE{2017MNRAS.466.3217Z,
       author = {{Zhang}, Kai and {Yan}, Renbin and {Bundy}, Kevin and {Bershady}, Matthew and {Haffner}, L. Matthew and {Walterbos}, Ren{\'e} and {Maiolino}, Roberto and {Tremonti}, Christy and {Thomas}, Daniel and {Drory}, Niv and {Jones}, Amy and {Belfiore}, Francesco and {S{\'a}nchez}, Sebastian F. and {Diamond-Stanic}, Aleksandar M. and {Bizyaev}, Dmitry and {Nitschelm}, Christian and {Andrews}, Brett and {Brinkmann}, Jon and {Brownstein}, Joel R. and {Cheung}, Edmond and {Li}, Cheng and {Law}, David R. and {Roman Lopes}, Alexandre and {Oravetz}, Daniel and {Pan}, Kaike and {Storchi Bergmann}, Thaisa and {Simmons}, Audrey},
        title = "{SDSS-IV MaNGA: the impact of diffuse ionized gas on emission-line ratios, interpretation of diagnostic diagrams and gas metallicity measurements}",
      journal = {\mnras},
         year = 2017,
        month = apr,
       volume = {466},
       number = {3},
        pages = {3217-3243},
          doi = {10.1093/mnras/stw3308},
archivePrefix = {arXiv},
       eprint = {1612.02000},
 primaryClass = {astro-ph.GA},
       adsurl = {https://ui.adsabs.harvard.edu/abs/2017MNRAS.466.3217Z}
}

@ARTICLE{2012A&A...546A...4T,
       author = {{Tamm}, A. and {Tempel}, E. and {Tenjes}, P. and {Tihhonova}, O. and {Tuvikene}, T.},
        title = "{Stellar mass map and dark matter distribution in M 31}",
      journal = {\aap},
         year = 2012,
        month = oct,
       volume = {546},
          eid = {A4},
        pages = {A4},
          doi = {10.1051/0004-6361/201220065},
archivePrefix = {arXiv},
       eprint = {1208.5712},
 primaryClass = {astro-ph.CO},
       adsurl = {https://ui.adsabs.harvard.edu/abs/2012A&A...546A...4T}
}

@ARTICLE{1997A&A...321..111P,
       author = {{Prugniel}, P. and {Simien}, F.},
        title = "{The fundamental plane of early-type galaxies: non-homology of the spatial structure.}",
      journal = {\aap},
         year = 1997,
        month = may,
       volume = {321},
        pages = {111-122},
       adsurl = {https://ui.adsabs.harvard.edu/abs/1997A&A...321..111P}
}

@ARTICLE{2025arXiv250801102G,
       author = {{Gunasekera}, Chamani M. and {van Hoof}, Peter A.~M. and {Dehghanian}, Maryam and {Chakraborty}, Priyanka and {Shaw}, Gargi and {Bianchi}, Stefano and {Chatzikos}, Marios and {Tsujimoto}, Masahiro and {Ferland}, Gary J.},
        title = "{The 2025 Release of Cloudy}",
      journal = {arXiv e-prints},
         year = 2025,
        month = aug,
          eid = {arXiv:2508.01102},
        pages = {arXiv:2508.01102},
          doi = {10.48550/arXiv.2508.01102},
archivePrefix = {arXiv},
       eprint = {2508.01102},
 primaryClass = {astro-ph.GA},
       adsurl = {https://ui.adsabs.harvard.edu/abs/2025arXiv250801102G}
}

@ARTICLE{2018MNRAS.478.5379D,
       author = {{Dong}, Hui and {Olsen}, Knut and {Lauer}, Tod and {Saha}, Abhijit and {Li}, Zhiyuan and {Garc{\'\i}a-Benito}, Ruben and {Sch{\"o}del}, Rainer},
        title = "{The star formation history in the M31 bulge}",
      journal = {\mnras},
         year = 2018,
        month = aug,
       volume = {478},
       number = {4},
        pages = {5379-5403},
          doi = {10.1093/mnras/sty1381},
archivePrefix = {arXiv},
       eprint = {1805.07225},
 primaryClass = {astro-ph.GA},
       adsurl = {https://ui.adsabs.harvard.edu/abs/2018MNRAS.478.5379D}
}

@ARTICLE{2017RMxAA..53..385F,
       author = {{Ferland}, G.~J. and {Chatzikos}, M. and {Guzm{\'a}n}, F. and {Lykins}, M.~L. and {van Hoof}, P.~A.~M. and {Williams}, R.~J.~R. and {Abel}, N.~P. and {Badnell}, N.~R. and {Keenan}, F.~P. and {Porter}, R.~L. and {Stancil}, P.~C.},
        title = "{The 2017 Release Cloudy}",
      journal = {\rmxaa},
         year = 2017,
        month = oct,
       volume = {53},
        pages = {385-438},
archivePrefix = {arXiv},
       eprint = {1705.10877},
 primaryClass = {astro-ph.GA},
       adsurl = {https://ui.adsabs.harvard.edu/abs/2017RMxAA..53..385F}
}

@ARTICLE{1980A&A....87..152H,
       author = {{Heckman}, T.~M.},
        title = "{An optical and radio survey of the nuclei of bright galaxies. Activity in normal galactic nuclei.}",
      journal = {\aap},
         year = 1980,
        month = jul,
       volume = {500},
        pages = {187-199},
       adsurl = {https://ui.adsabs.harvard.edu/abs/1980A&A....87..152H}
}

@ARTICLE{1999ApJ...516..672H,
       author = {{Ho}, Luis C.},
        title = "{The Spectral Energy Distributions of Low-Luminosity Active Galactic Nuclei}",
      journal = {\apj},
         year = 1999,
        month = may,
       volume = {516},
       number = {2},
        pages = {672-682},
          doi = {10.1086/307137},
archivePrefix = {arXiv},
       eprint = {astro-ph/9905012},
 primaryClass = {astro-ph},
       adsurl = {https://ui.adsabs.harvard.edu/abs/1999ApJ...516..672H}
}

@ARTICLE{2006MNRAS.372..961K,
       author = {{Kewley}, Lisa J. and {Groves}, Brent and {Kauffmann}, Guinevere and {Heckman}, Tim},
        title = "{The host galaxies and classification of active galactic nuclei}",
      journal = {\mnras},
         year = 2006,
        month = nov,
       volume = {372},
       number = {3},
        pages = {961-976},
          doi = {10.1111/j.1365-2966.2006.10859.x},
archivePrefix = {arXiv},
       eprint = {astro-ph/0605681},
 primaryClass = {astro-ph},
       adsurl = {https://ui.adsabs.harvard.edu/abs/2006MNRAS.372..961K}
}

@ARTICLE{2019ARA&A..57..511K,
       author = {{Kewley}, Lisa J. and {Nicholls}, David C. and {Sutherland}, Ralph S.},
        title = "{Understanding Galaxy Evolution Through Emission Lines}",
      journal = {\araa},
         year = 2019,
        month = aug,
       volume = {57},
        pages = {511-570},
          doi = {10.1146/annurev-astro-081817-051832},
archivePrefix = {arXiv},
       eprint = {1910.09730},
 primaryClass = {astro-ph.GA},
       adsurl = {https://ui.adsabs.harvard.edu/abs/2019ARA&A..57..511K}
}

@ARTICLE{2000AJ....119.2745K,
       author = {{Kong}, Xu and {Zhou}, Xu and {Chen}, Jiansheng and {Cheng}, Fuzhen and {Jiang}, Zhaoji and {Zhu}, Jin and {Zheng}, Zhongyuan and {Mao}, Shude and {Shang}, Zhaohui and {Fan}, Xiaohui and {Byun}, Yong-Ik and {Chen}, Rui and {Chen}, Wen-ping and {Deng}, Licai and {Hester}, J. Jeff and {Li}, Yong and {Lin}, Weipeng and {Su}, Hongjun and {Sun}, Wei-hsin and {Tsay}, Wean-shun and {Windhorst}, Rogier A. and {Wu}, Hong and {Xia}, Xiaoyang and {Xu}, Wen and {Xue}, Suijian and {Yan}, Haojing and {Zheng}, Zheng and {Zou}, Zhenglong},
        title = "{Spatially Resolved Spectrophotometry of M81: Age, Metallicity, and Reddening Maps}",
      journal = {\aj},
         year = 2000,
        month = jun,
       volume = {119},
       number = {6},
        pages = {2745-2756},
          doi = {10.1086/301396},
archivePrefix = {arXiv},
       eprint = {astro-ph/0002266},
 primaryClass = {astro-ph},
       adsurl = {https://ui.adsabs.harvard.edu/abs/2000AJ....119.2745K}
}

@ARTICLE{2003AJ....125.1226D,
       author = {{Devereux}, Nick and {Ford}, Holland and {Tsvetanov}, Zlatan and {Jacoby}, George},
        title = "{STIS Spectroscopy of the Central 10 Parsecs of M81: Evidence for a Massive Black Hole}",
      journal = {\aj},
         year = 2003,
        month = mar,
       volume = {125},
       number = {3},
        pages = {1226-1235},
          doi = {10.1086/367595},
       adsurl = {https://ui.adsabs.harvard.edu/abs/2003AJ....125.1226D}
}

@ARTICLE{2003MNRAS.342..345C,
       author = {{Cappellari}, Michele and {Copin}, Yannick},
        title = "{Adaptive spatial binning of integral-field spectroscopic data using Voronoi tessellations}",
      journal = {\mnras},
         year = 2003,
        month = jun,
       volume = {342},
       number = {2},
        pages = {345-354},
          doi = {10.1046/j.1365-8711.2003.06541.x},
archivePrefix = {arXiv},
       eprint = {astro-ph/0302262},
 primaryClass = {astro-ph},
       adsurl = {https://ui.adsabs.harvard.edu/abs/2003MNRAS.342..345C}
}

@ARTICLE{2011ApJ...728L..10L,
       author = {{Li}, Zhiyuan and {Garcia}, Michael R. and {Forman}, William R. and {Jones}, Christine and {Kraft}, Ralph P. and {Lal}, Dharam V. and {Murray}, Stephen S. and {Wang}, Q. Daniel},
        title = "{The Murmur of the Hidden Monster: Chandra's Decadal View of the Supermassive Black Hole in M31}",
      journal = {\apjl},
         year = 2011,
        month = feb,
       volume = {728},
       number = {1},
          eid = {L10},
        pages = {L10},
          doi = {10.1088/2041-8205/728/1/L10},
archivePrefix = {arXiv},
       eprint = {1011.1224},
 primaryClass = {astro-ph.HE},
       adsurl = {https://ui.adsabs.harvard.edu/abs/2011ApJ...728L..10L}
}

@ARTICLE{1992ApJ...397L..79F,
       author = {{Filippenko}, Alexei V. and {Terlevich}, Roberto},
        title = "{O-Star Photoionization Models of Liners with Weak [O i] lambda 6300 Emission}",
      journal = {\apjl},
         year = 1992,
        month = oct,
       volume = {397},
        pages = {L79},
          doi = {10.1086/186549},
       adsurl = {https://ui.adsabs.harvard.edu/abs/1992ApJ...397L..79F}
}

@ARTICLE{2014ApJ...780..172D,
       author = {{Draine}, B.~T. and {Aniano}, G. and {Krause}, Oliver and {Groves}, Brent and {Sandstrom}, Karin and {Braun}, Robert and {Leroy}, Adam and {Klaas}, Ulrich and {Linz}, Hendrik and {Rix}, Hans-Walter and {Schinnerer}, Eva and {Schmiedeke}, Anika and {Walter}, Fabian},
        title = "{Andromeda's Dust}",
      journal = {\apj},
         year = 2014,
        month = jan,
       volume = {780},
       number = {2},
          eid = {172},
        pages = {172},
          doi = {10.1088/0004-637X/780/2/172},
archivePrefix = {arXiv},
       eprint = {1306.2304},
 primaryClass = {astro-ph.CO},
       adsurl = {https://ui.adsabs.harvard.edu/abs/2014ApJ...780..172D}
}

@ARTICLE{1994A&A...292...13B,
       author = {{Binette}, L. and {Magris}, C.~G. and {Stasi{\'n}ska}, G. and {Bruzual}, A.~G.},
        title = "{Photoionization in elliptical galaxies by old stars.}",
      journal = {\aap},
         year = 1994,
        month = dec,
       volume = {292},
        pages = {13-19},
       adsurl = {https://ui.adsabs.harvard.edu/abs/1994A&A...292...13B}
}

@ARTICLE{2014ApJ...780..128I,
       author = {{Ibata}, Rodrigo A. and {Lewis}, Geraint F. and {McConnachie}, Alan W. and {Martin}, Nicolas F. and {Irwin}, Michael J. and {Ferguson}, Annette M.~N. and {Babul}, Arif and {Bernard}, Edouard J. and {Chapman}, Scott C. and {Collins}, Michelle and {Fardal}, Mark and {Mackey}, A.~D. and {Navarro}, Julio and {Pe{\~n}arrubia}, Jorge and {Rich}, R. Michael and {Tanvir}, Nial and {Widrow}, Lawrence},
        title = "{The Large-scale Structure of the Halo of the Andromeda Galaxy. I. Global Stellar Density, Morphology and Metallicity Properties}",
      journal = {\apj},
         year = 2014,
        month = jan,
       volume = {780},
       number = {2},
          eid = {128},
        pages = {128},
          doi = {10.1088/0004-637X/780/2/128},
archivePrefix = {arXiv},
       eprint = {1311.5888},
 primaryClass = {astro-ph.GA},
       adsurl = {https://ui.adsabs.harvard.edu/abs/2014ApJ...780..128I}
}

@INPROCEEDINGS{1998tbha.conf..148N,
       author = {{Narayan}, R. and {Mahadevan}, R. and {Quataert}, E.},
        title = "{Advection-dominated accretion around black holes}",
    booktitle = {Theory of Black Hole Accretion Disks},
         year = 1998,
       editor = {{Abramowicz}, M.~A. and {Bj{\"o}rnsson}, G. and {Pringle}, J.~E.},
        month = jan,
        pages = {148-182},
          doi = {10.48550/arXiv.astro-ph/9803141},
archivePrefix = {arXiv},
       eprint = {astro-ph/9803141},
 primaryClass = {astro-ph},
       adsurl = {https://ui.adsabs.harvard.edu/abs/1998tbha.conf..148N}
}

@ARTICLE{2018A&A...618A.156S,
       author = {{Saglia}, R.~P. and {Opitsch}, M. and {Fabricius}, M.~H. and {Bender}, R. and {Bla{\~n}a}, M. and {Gerhard}, O.},
        title = "{Stellar populations of the central region of M 31}",
      journal = {\aap},
         year = 2018,
        month = oct,
       volume = {618},
          eid = {A156},
        pages = {A156},
          doi = {10.1051/0004-6361/201732517},
archivePrefix = {arXiv},
       eprint = {1807.09284},
 primaryClass = {astro-ph.GA},
       adsurl = {https://ui.adsabs.harvard.edu/abs/2018A&A...618A.156S}
}

@ARTICLE{1998ApJ...504..113B,
       author = {{Brown}, Thomas M. and {Ferguson}, Henry C. and {Stanford}, S.~A. and {Deharveng}, Jean-Michel},
        title = "{Color-Luminosity Relations for the Resolved Hot Stellar Populations in the Centers of M31 and M32}",
      journal = {\apj},
         year = 1998,
        month = sep,
       volume = {504},
       number = {1},
        pages = {113-138},
          doi = {10.1086/306079},
archivePrefix = {arXiv},
       eprint = {astro-ph/9803327},
 primaryClass = {astro-ph},
       adsurl = {https://ui.adsabs.harvard.edu/abs/1998ApJ...504..113B}
}

@ARTICLE{2005ApJ...631..280B,
       author = {{Bender}, Ralf and {Kormendy}, John and {Bower}, Gary and {Green}, Richard and {Thomas}, Jens and {Danks}, Anthony C. and {Gull}, Theodore and {Hutchings}, J.~B. and {Joseph}, C.~L. and {Kaiser}, M.~E. and {Lauer}, Tod R. and {Nelson}, Charles H. and {Richstone}, Douglas and {Weistrop}, Donna and {Woodgate}, Bruce},
        title = "{HST STIS Spectroscopy of the Triple Nucleus of M31: Two Nested Disks in Keplerian Rotation around a Supermassive Black Hole}",
      journal = {\apj},
         year = 2005,
        month = sep,
       volume = {631},
       number = {1},
        pages = {280-300},
          doi = {10.1086/432434},
archivePrefix = {arXiv},
       eprint = {astro-ph/0509839},
 primaryClass = {astro-ph},
       adsurl = {https://ui.adsabs.harvard.edu/abs/2005ApJ...631..280B}
}

@ARTICLE{1988AJ.....95..438C,
       author = {{Ciardullo}, Robin and {Rubin}, Vera C. and {Ford}, Jr., W. Kent and {Jacoby}, George H. and {Ford}, Holland C.},
        title = "{The Morphology of Ionized Gas in M31's Bulge}",
      journal = {\aj},
         year = 1988,
        month = feb,
       volume = {95},
        pages = {438},
          doi = {10.1086/114645},
       adsurl = {https://ui.adsabs.harvard.edu/abs/1988AJ.....95..438C}
}
\bibliographystyle{aasjournal}



\end{document}